# TotalSegmentator: Robust Segmentation of 104 Anatomical Structures in CT images

Jakob Wasserthal PhD[1], Hanns-Christian Breit MD[1], Manfred T. Meyer MD[1], Maurice Pradella MD[1], Daniel Hinck[1], Alexander W. Sauter MD[1], Tobias Heye MD[1], Daniel Boll MD[1], Joshy Cyriac MSc[1], Shan Yang PhD[1], Michael Bach PhD[1], Martin Segeroth MD[1]

[1] Clinic of Radiology and Nuclear Medicine, University Hospital Basel, Basel, Switzerland

Correspondence to jakob.wasserthal@usb.ch

## Abstract

**Purpose**

To present a deep learning segmentation model that can automatically and robustly segment all major anatomical structures in body CT images.

**Materials and Methods**

In this retrospective study, 1204 CT examinations (from the years 2012, 2016, and 2020) were used to segment 104 anatomical structures (27 organs, 59 bones, 10 muscles, 8 vessels) relevant for use cases such as organ volumetry, disease characterization, and surgical or radiotherapy planning. The CT images were randomly sampled from routine clinical studies and thus represent a real-world dataset (different ages, pathologies, scanners, body parts, sequences, and sites). The authors trained an nnU-Net segmentation algorithm on this dataset and calculated Dice similarity coefficients (Dice) to evaluate the model's performance. The trained algorithm was applied to a second dataset of 4004 whole-body CT examinations to investigate age dependent volume and attenuation changes.

**Results**

The proposed model showed a high Dice score (0.943) on the test set, which included a wide range of clinical data with major pathologies. The model significantly outperformed another publicly available segmentation model on a separate dataset (Dice score, 0.932 versus 0.871, respectively; $p<0.001$). The aging study demonstrated significant correlations between age and volume and mean attenuation for a variety of organ groups (e.g., age and aortic volume [$r_s = 0.64$; $p<0.001$]; age and mean attenuation of the autochthonous dorsal musculature [$r_s = −0.74$; $p<0.001$]).

**Conclusion**

The developed model enables robust and accurate segmentation of 104 anatomical structures. The annotated dataset (https://doi.org/10.5281/zenodo.6802613) and toolkit (https://www.github.com/wasserth/TotalSegmentator) are publicly available.

# 1. Introduction

In the last few years, both the number of CT examinations performed and the available computing power have steadily increased (1,2). Moreover, the capability of image analysis algorithms has vastly improved given advances in deep learning techniques (3,4). The resulting increases in data, computational power, and algorithm quality have enabled radiologic studies using large sample sizes. For many of these studies, segmentation of anatomical structures plays an important role. Segmentation is useful for extracting advanced biomarkers based on radiologic images, automatically detecting pathologies, or quantifying tumor load (5). In routine clinical analysis, segmentation is already used for applications such as surgical and radiotherapy planning (6). Thus, the associated algorithms could ultimately enter routine clinical use to improve the quality of radiologic reports and reduce radiologist workload.

For most applications, segmentation of the relevant anatomical structure is the first step. Building and training a segmentation algorithm, however, is complex, as it requires tedious manual annotation of training data and technical expertise for training the algorithm. Providing a ready-to-use segmentation toolkit that enables automatic segmentation of most of the major anatomical structures in CT images would considerably simplify many radiology studies, thereby accelerating research in the field.

Several publicly available segmentation models are currently available. However, these models are generally specific for a single organ (e.g., the pancreas, spleen, colon, or lung) (7–11). Additionally, the models cover only a small subset of relevant anatomical structures and are trained on relatively small datasets that are not representative of routine clinical imaging, which is characterized by differences in contrast phases, acquisition settings, and diverse pathologies. Thus, researchers must often build and train their own segmentation models, which can be costly.

To overcome this problem, we aimed to develop a model with the following characteristics: (1) publicly available (including its training data), (2) easy to use, (3) segments most anatomically relevant structures throughout the body, and (4) exhibits robust performance in any clinical setting. As an example application, we applied the segmentation tool to a large dataset of 4004 patients with whole-body CT scans collected in a polytrauma setting and analyzed age dependent changes of the volume and attenuation of different structures.

# 2. Materials and Methods

The ethics waiver for this retrospective study was approved by the Ethics Committee Northwest and Central Switzerland (EKNZ BASEC Req-2022-00495).

## 2.1. Datasets

Two datasets were aggregated for this study: one dataset for training the proposed model ("training dataset") and a second dataset for the aging study example application ("aging study dataset").

**Training dataset**

To generate a comprehensive and highly variant dataset, 1368 CT images were randomly sampled from the years 2012, 2016, and 2020 from the University Hospital Basel picture archiving and communication system (PACS). CT series of upper and lower extremities (n = 37), CT series with missing sections (n = 87), and CT series for which the human annotator was unable to segment certain structures because of high ambiguity (e.g., structures highly distorted due to pathology) (n = 40) were excluded. The CT series were sampled randomly from each examination to obtain high variety of data. All images were resampled to 1.5 mm isotropic resolution. The final dataset of 1204 CT series was divided into a training dataset of 1082 patients (90%), a validation dataset of 57 patients (5%), and a test dataset of 65 patients (5%) (Figure 1a).

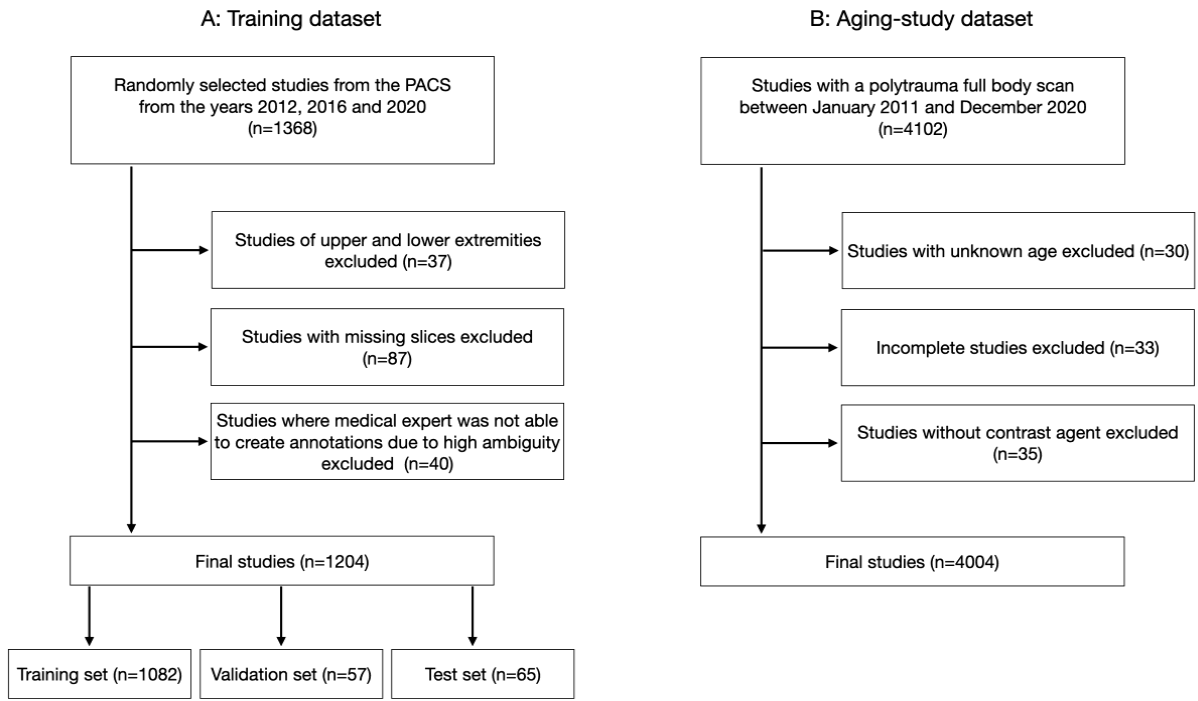

Figure 1a: Diagram showing the inclusion of patients into the study

**Aging-study dataset**

All patients with polytrauma who received whole-body CT scans between 2011 and 2020 at the University Hospital Basel were initially included (n = 4102). Patients with unknown age (n = 30), incomplete images (n = 33), or who underwent studies without administration of contrast agent (n = 35) were excluded. In all patients, the same examination protocol was applied: contrast-enhanced, whole-body CT in an arteriovenous split-bolus phase, with a similar amount of contrast agent. Thus, we assumed that variation of attenuation (in Hounsfield units [HU]) due to the contrast agent was minimal for all images and that the attenuation values were comparable.

*2.2. Data Annotation*

We identified 104 anatomical structures for segmentation (Figure 2; Supplemental Materials S1). The Nora Imaging Platform was used for manual segmentation or further refinement of generated segmentations (12). Segmentation was supervised by two physicians with 3 (M.S.) and 6 years (H.B.) of experience in body imaging, respectively. The work was split between them.

If an existing model for a given structure was publicly available (Supplemental Materials S2), that model was used to create a first segmentation, which was then validated and refined manually (10,13–16)

To speed the process further, we used an iterative learning approach, as follows. After manual segmentation of the first 5 patients was completed, a preliminary nnU-Net was trained, and its predictions were manually refined, if necessary. Retraining of the nnU-Net was performed after reviewing and refining 5 patients, 20 patients, and 100 patients (Figure 1b).

In the end, all 1204 CT examinations had annotations that were manually reviewed and corrected whenever necessary. These final annotations served as the ground truth for training and testing. The model was trained on the dataset of 1082 patients, validated on the dataset of 57 patients and tested on the dataset of 65 patients. This final model was independent of the intermediate models trained during the annotation workflow, which reduced bias in the test set to a minimum. Using completely manual annotations in the test set would have introduced a distribution shift and thus greater bias.

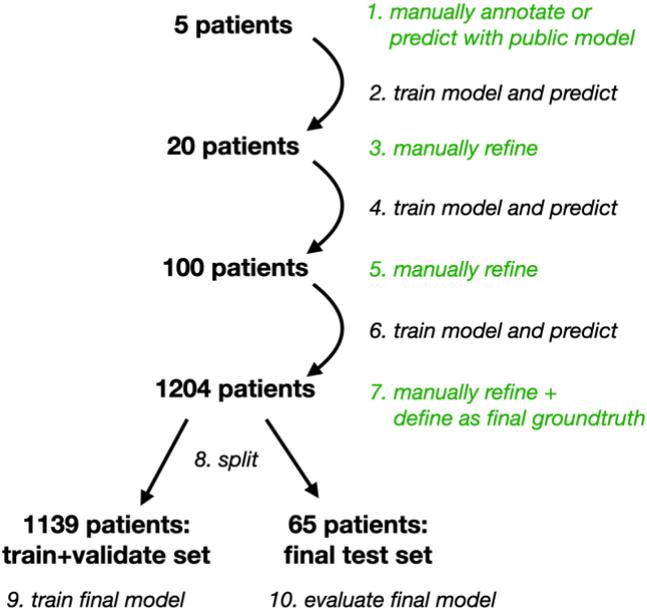

Figure 1b: Diagram showing the iterative annotation workflow of the training dataset. Steps involving manual annotation are shown in green. In step 9, a completely new model was trained independently of the intermediate models (step 2,4,6). This avoids leakage of information from the test set into the training set. PACS = picture archiving and communication system

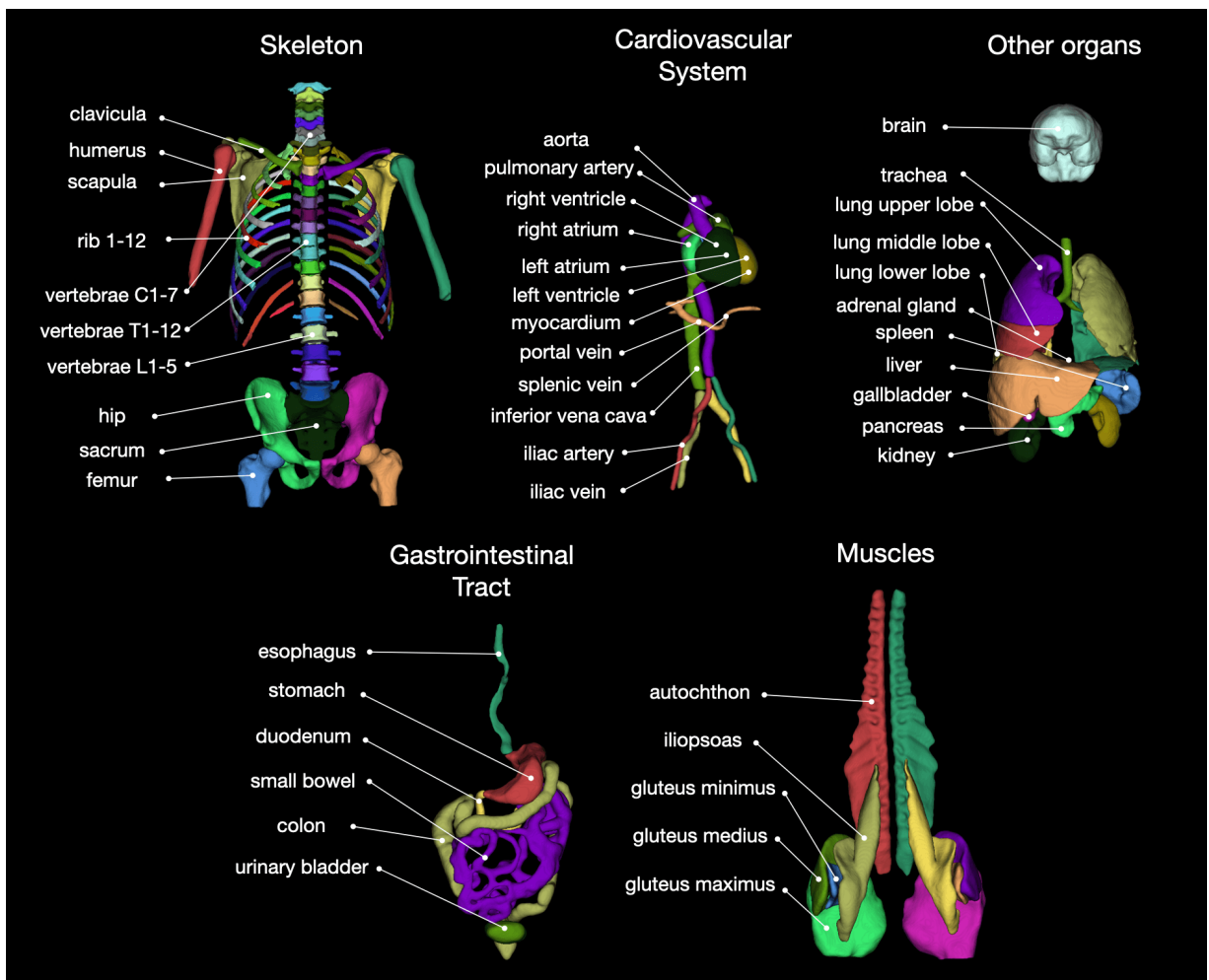

Figure 2: Overview of all 104 anatomical structures segmented by the TotalSegmentator.

## 2.3. Model

We used the model from the nnU-Net framework, which is a U-Net–based implementation that automatically configures all hyperparameters based on the dataset characteristics (17,18). One model was trained on CT scans with 1.5 mm isotropic resolution. To allow for lower technical requirements (random-access memory [RAM] and graphics processing unit [GPU] memory), we also trained a second model on 3 mm isotropic resolution (for more details on the training, see Supplemental Materials S4). The runtime for the prediction of one case was measured on a local workstation with an Intel Core i9 3.5GHz CPU and Nvidia GeForce RTX 3090 GPU.

## 2.4. Statistical Analysis

**Training dataset**

As evaluation metrics, the Dice similarity coefficient (Dice), a commonly used spatial overlap index, and the normalized surface distance (NSD), which measures how often the surface distance is <3 mm, were calculated between the predicted segmentations and the human approved ground truth segmentations. Both metrics range between 0 (worst) and 1 (best) and were calculated on the test set.

For additional evaluation, we compared our model to a nnU-Net trained on the dataset from the "Multi-Atlas Labeling Beyond the Cranial Vault Challenge"

([https://www.synapse.org/#!Synapse:syn3193805/wiki/217780](https://www.synapse.org/#!Synapse:syn3193805/wiki/217780)) (BTCV dataset) acquired at Vanderbilt University Medical Center. As that dataset provided labels for only 13 structures (Supplemental Materials S5), the comparison was limited to those 13 structures. We ran two comparisons, one on our test set and one on the BTCV dataset. The 95% confidence intervals (CIs) were calculated using nonparametric percentile bootstrapping with 10000 iterations. Wilcoxon signed rank test was used to compare the Dice and NSD metrics between our model and the BTCV model. P-values <0.05 were considered significant.

**Aging study dataset**
To evaluate the effect of age on different structures, the correlation between age and volume and the mean attenuation in HU were calculated for all structures, excluding structures with failed segmentations. The segmentation of a body structure was assumed as failed if the volume of the respective body structure was too small to be anatomical plausible given by a lower bound (Supplemental Materials Table S1). The Kolmogorov-Smirnov test was used to evaluate whether continuous variables were normally distributed. The association between continuous variables was examined using Spearman's rank correlation coefficient. Patients were grouped into four age quartiles and compared using Kruskal Wallis test. Post-hoc analysis was performed using Wilcoxon rank sum test. Bonferroni correction was performed and p-values of <0.0001 were considered significant. Outliers are not shown in the figures to maintain scaling.

## 3. Results

### 3.1. Characteristics of the Study Sample

Data regarding basic demographic characteristics of patients included in the training dataset of 1204 CT images are shown in Figure 3. The dataset contained a high variety of CT images, with differences in slice thickness, resolution, and contrast phase (native, arterial, portal venous, late phase, and others). Dual-energy CT images obtained using different tube voltages were also included. Different kernels (soft tissue kernel, bone kernel), as well as CT images from 8 different sites and 16 different scanners were included in the dataset; however, most images were acquired using a Siemens manufacturer. A total of 404 patients showed no signs of pathology, whereas 645 showed different types of pathology (tumor, vascular, trauma, inflammation, bleeding, other). Information regarding presence of pathologies was not available for 155 patients due to missing radiologic reports (Figure 3).

The aging-study dataset of CT scans from 4004 patients showed uniform age distribution, ranging from 18 to 100 years (Figure S2). The sex distribution was less balanced (2543 (63.5%) males, 1461 (36.5%) females).

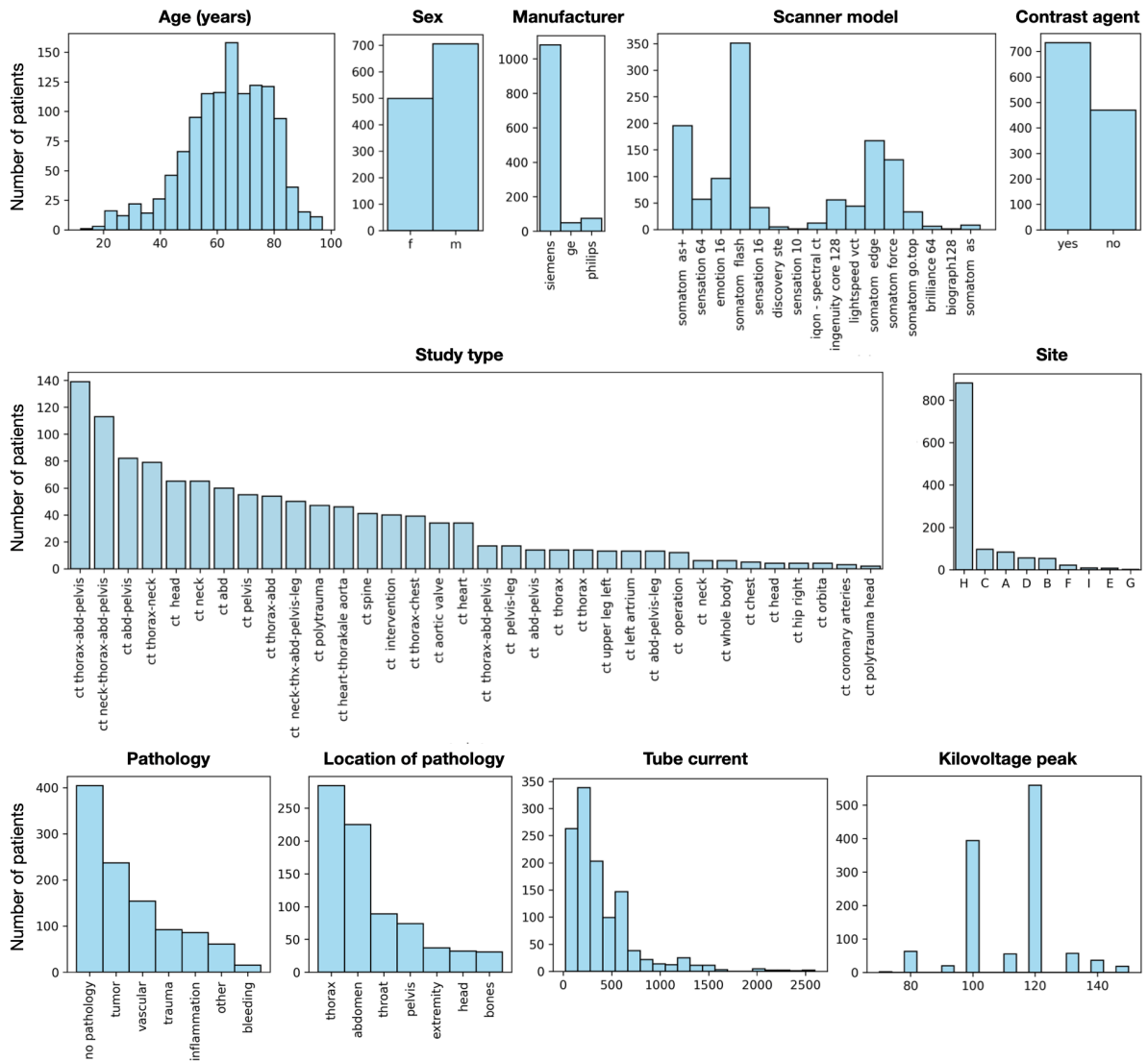

Figure 3: Graphs showing the distribution of different parameters of the training dataset, demonstrating the dataset's high diversity.

*3.4. Segmentation Evaluation*

The model trained on CT images with a slice thickness of 1.5 mm showed high accuracy. The Dice score was 0.943 (95% CI [0.938, 0.947]), and the NSD was 0.966 (95% CI [0.962, 0.971]). The 3 mm model showed a lower Dice score of 0.840 (95% CI [0.836, 0.844]), but the NSD was 0.966 (95% CI [0.962, 0.969]), as minor inaccuracies introduced by the lower resolution were still within the bounds of the 3 mm distance threshold. Thus, the 3 mm model still delivered correct results, but the borders were less precise. Results for each structure independently are shown in Figure S1 and at https://github.com/wasserth/TotalSegmentator/blob/master/resources/results_all_classes.json.

In a direct comparison of our 1.5 mm model to an nnU-Net trained on the BTCV dataset, our odel achieved a significantly higher Dice (0.932, 95% CI [0.920, 0.942]) vs 0.871 (95% CI [0.855, 0.887]), respectively; p<0.001) and NSD score (0.971, 95% CI [0.961, 0.979]) vs 0.921 (95% CI [0.907, 0.936]); p<0.001 on our test set. When testing our 1.5 mm model on the BTCV dataset, it achieved a Dice score of 0.849 (95% CI [0.833, 0.862]) and NSD score of 0.932 (95% CI [0.920, 0.943]), showing generalizability to CT studies from a different continent. Our model achieved higher values (p<0.001) than the nnU-Net trained on the BTCV dataset itself (Dice, 0.839, 95% CI [0.821, 0.856]; NSD, 0.915, 95% CI [0.900, 0.930]) (for more information, see Supplemental Materials S7).

*Typical Failure Cases*

Despite the high Dice score and NSD, our model failed in some cases. Figure 4 shows the most typical failure cases, such as missing small parts of the colon or iliac arteries and mixing up neighboring vertebrae and ribs.

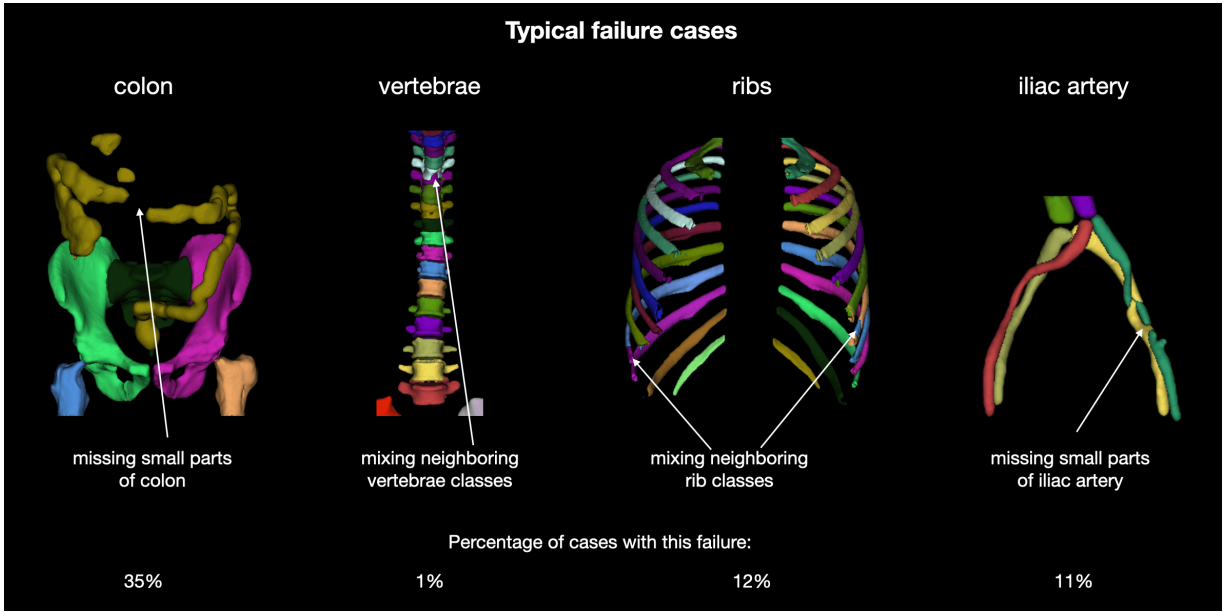

Figure 4: Overview of typical failure cases of the proposed model. Users should be aware that these problems may occur.

*Performance on Pathologic Cases*

As our model was trained on a diverse dataset, it generated robust results on patients with major pathologies. Figure 5 shows qualitative results for several pathologies.

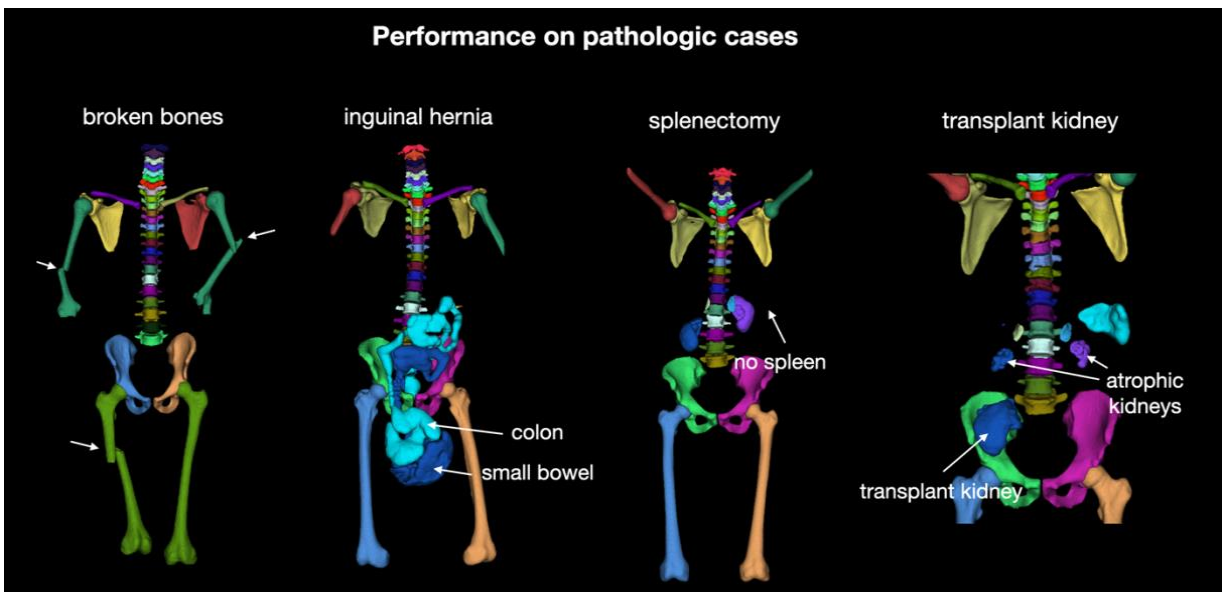

Figure 5: Overview of performance of the proposed model on different pathologies on the test set. Our model showed robust, accurate results even when structures were distorted (broken bones), displaced (bowels displaced by inguinal hernia), completely missing (splenectomy) or duplicated (transplant kidney).

*Runtime*

Table 1 shows an overview of the runtime, RAM requirements, and GPU memory requirements of the high-resolution (1.5 mm) and low-resolution (3 mm) models for three CT studies with different dimensions: a small study of an abdomen with matrix size of 512 × 512 × 280 voxels, a medium-sized study of thorax and abdomen with matrix size of 512 × 512 × 458 voxels, and a large study from head to knee with matrix size of 512 × 512 × 824 voxels.

Table 1: Overview of Runtime, RAM Requirements and GPU Memory Requirements of the High Resolution (1.5 mm) and Low Resolution (3 mm) Models on Three Different-sized CT Studies

|  | 1.5mm model | | | 3mm model | | |
| --- | --- | --- | --- | --- | --- | --- |
| Study size (in voxels) | Runtime | RAM | GPU Mem | Runtime | RAM | GPU Mem |
| Small (512 x 512 x 280) | 1min 17s | 7.6GB | 6.1GB | 34s | 7.4GB | 5.2GB |
| Medium (512 x 512 x 458) | 2min 49s | 10.6GB | 8.5GB | 53s | 8.4GB | 7.4GB |
| Large (512 x 512 x 824) | 3min 32s | 11.8GB | 11.4GB | 1min 23s | 10.6GB | 7.5GB |

Note.—The three CT studies were as follows: one small study of only abdomen with matrix size of 512 x 512 x 280 voxels, one medium study of the thorax and abdomen with matrix size of 512 x 512 x 458 voxels, and one large study from head to knee with matrix size of 512 x 512 x 824 voxels. The runtime was measured on a local workstation with an Intel Core i9 3.5GHz CPU and a Nvidia GeForce RTX 3090 GPU. GPU = graphics processing unit, Mem = memory, RAM = random-access memory

*Evaluation of Age-related Differences*

For the aging-study dataset, we observed a negative correlation between age and attenuation in the clavicula ($r_s = -0.53$; p<0.0001), hips ($r_s = -0.61$; p<0.0001, Figure 6), and all ribs and the scapula (Supplemental Materials Figure S3). A moderately negative correlation was observed between age and mean attenuation of the lumbar vertebrae (lumbar vertebra 4, $r_s = -0.55$; p<0.0001).

Negative correlations between age and attenuation were also observed for the musculature, with moderate to strong correlations for the autochthonous dorsal musculature ($r_s = -0.74$; p<0.0001) and gluteal musculature (gluteus maximus: $r_s = -0.51$; p<0.0001, gluteus medius: $r_s = -0.61$; p<0.0001, gluteus minimus: $r_s = -0.79$; p<0.0001). Age was moderately and strongly negatively correlated with CT attenuation and volume of the iliopsoas muscle, respectively ($r_s = -0.57$; p<0.0001 and $r_s = -0.61$; p<0.0001, Figure 6).

A positive correlation was observed between the volume of the aorta and patient age ($r_s = 0.64$; p<0.0001, Figure 6), most likely due to aneurysm development. The correlation was less positive for the iliac arteries ($r_s = 0.33$; p<0.0001).

Regarding organ volumetry, age was negatively correlated to kidney ($r_s = -0.49$; p<0.0001) and pancreas volume ($r_s = -0.49$; p<0.0001).

The same analysis was done for all 104 classes (Supplement Materials Figure S3).

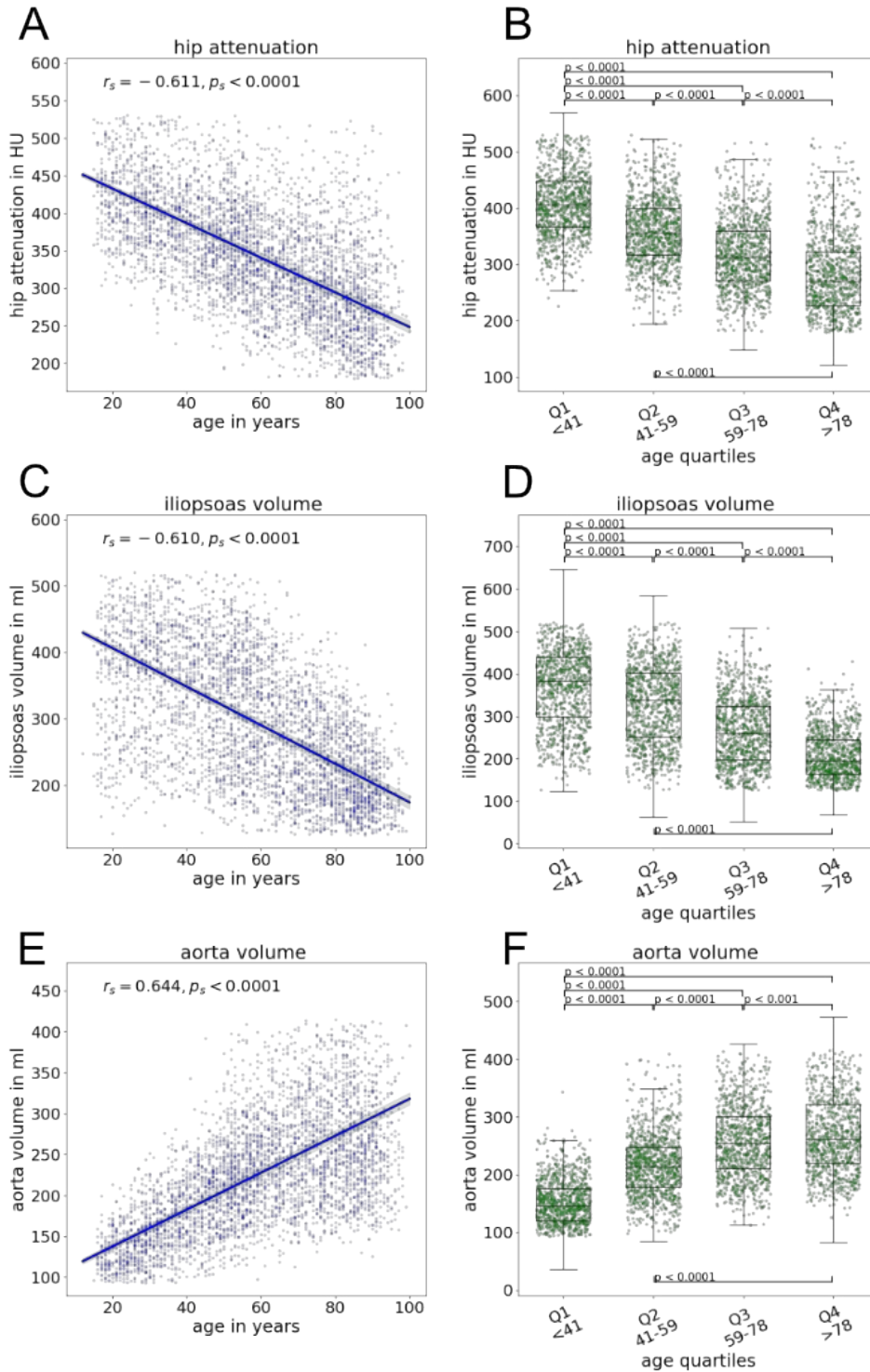

Figure 6: Example correlations of CT attenuation and volume with patient age. A) Graph shows negative correlation between hip attenuation and patient age, B) Boxplots of hip attentuation for age quartiles show a decrease with increasing age, C) Graph shows negative correlation between iliopsoas muscle volume and patient age, D) Boxplots of iliopsoas muscle volume for age quartiles show a decrease with increasing age, E) Graph shows positive correlation between aortic volume and patient age, F) Boxplots of aortic volume for age quartiles show an increase with increasing age.

# 4. Discussion

In this study, we developed a tool for segmentation of 104 anatomical structures on 1204 CT datasets obtained using different CT scanners, acquisition settings, and contrast phases. The tool demonstrated high accuracy (Dice score 0.943) and works robustly on a wide range of clinical data, outperforming other freely available segmentation tools. Furthermore, we evaluated and reported age-related changes in volume and attenuation in multiple organs using a large dataset of more than 4000 CT examinations.

Numerous models are available for segmentation of single or several organs on CT images (e.g., the pancreas, spleen, colon or lung), and work has also been conducted on segmenting several anatomical structures in one dataset and model (7–11). All previous models cover only a small subset of relevant anatomical structures and are trained on small datasets that are not representative of routine clinical imaging, which involves different contrast phases, acquisition settings, and diverse pathologies.(19)

To our knowledge, only three other studies have examined segmenting a larger number of structures in CT images. First, the algorithm reported by Chen et al. (20) segments 50 different structures. Apart from the fact that many organs and anatomical structures are still not represented in that model, neither the dataset nor the model are publicly available, and the dataset is relatively homogeneous (most of the training data came from the same scanner using the same CT sequence). Second, the algorithm developed by Sundar et al. segments 120 structures, and their model is publicly available (21). However, the model requires 256 GB of RAM, making it difficult to apply. Moreover, the training data consist of fewer than 100 individuals, making the model less robust for broad application to any CT data. Third, the algorithm developed by Trägårdh et al. segments 100 structures (22). However, as the 339 training samples are quite homogeneous, the model does not perform well on images with diverse slice or body orientations or involving different contrast phases.

Many segmentation models and datasets are not publicly available, which strongly reduces their benefit to the scientific community (6,23–25). Datasets that are made available often require time-consuming paperwork to request access (e.g., UK Biobank, NIH NDA) or are uploaded to data providers that are either difficult to use (e.g., The Cancer Imaging Archive (TCIA), which requires a third-party download manager) or rate limited (e.g., Google drive) Klicken oder tippen Sie hier, um Text einzugeben.. We made our model easily accessible by providing it as a pretrained python package. Our model requires less than 12 GB of RAM and does not require a GPU. Thus, it can be run on a normal laptop. In addition, our dataset is freely available to download; it does not require any access requests and can be downloaded with one click.

A nnU-Net based model was used for the present study, as it was shown to deliver accurate results across a wide range of tasks and has been established as the standard for medical image segmentation, outperforming most other methods (19). It might be possible to improve upon the default nnU-Net through more hyperparameter optimization and exploration of newer models, such as transformers (27).

Our model has multiple potential applications. Besides its use for surgery, rapid and readily available organ segmentation also allows for individual dosimetry, as shown for the liver and kidneys (20). Automated segmentation may also enhance research and provide normal or even age-dependent values (HU, volume, etc.) and biomarkers for clinicians. Combined with a lesion-detection model, our model could be used to estimate body part–specific tumor load. Moreover, our model can be used as a first step in building models to detect specific pathologies. Over 4500 researchers have already downloaded our model (https://zenodo.org/record/6802342), using it for a wide range of applications.

With our aging study, we demonstrated a sample application for our comprehensive segmentation model that could provide insights into the age dependency of organ volumes and attenuation. Such big-data evaluations were previously not feasible or required substantial time by expert researchers. Using a dataset of over 4000 patients who underwent a CT polytrauma scan, we showed correlations between age and volume of many segmented organs. Common literature values for normal organ sizes and age-dependent organ development are typically based on sample sizes of a few hundred patients. The number of diverse application examples, such as the evaluation of organ size or density as a function of age, sex, ethnicity, disease, medication intake, or drug use, is almost limitless and can provide a new (radiologic) approach for evaluating the physiology of disease processes.

A limitation of our study is that male patients were overrepresented in the study datasets, possibly because more males are part of the overall hospital population(28). We consider our model to serve as the basis for large radiologic population studies. For example, the model can be used to obtain new reference values for organ volumes or to create a new approach for evaluating different diseases using segmentation. In future research, we plan to add more anatomical structures to our dataset and model. Furthermore, we are preparing a more detailed aging study by using more patients, correcting for confounders, and analyzing more correlations.

In conclusion, we developed a CT segmentation model that is (1) publicly available (https://github.com/wasserth/TotalSegmentator), including training data (https://doi.org/10.5281/zenodo.6802613); (2) easy to use; (3) segments most anatomically relevant structures in the whole body; and (4) works robustly in any clinical setting.

## Supplemental Materials

### S1: List of all segmented structures

Spleen, kidney right, kidney left, gallbladder, liver, stomach, aorta, inferior vena cava, portal vein and splenic vein, pancreas, adrenal gland right, adrenal gland left, lung upper lobe left, lung lower lobe left, lung upper lobe right, lung middle lobe right, lung lower lobe right, vertebrae L5, vertebrae L4, vertebrae L3, vertebrae L2, vertebrae L1, vertebrae T12, vertebrae T11, vertebrae T10, vertebrae T9, vertebrae T8, vertebrae T7, vertebrae T6, vertebrae T5, vertebrae T4, vertebrae T3, vertebrae T2, vertebrae T1, vertebrae C7, vertebrae C6, vertebrae C5, vertebrae C4, vertebrae C3, vertebrae C2, vertebrae C1, esophagus, trachea, heart myocardium, heart atrium left, heart ventricle left, heart atrium right, heart ventricle right, pulmonary artery, brain, iliac artery left, iliac artery right, iliac vein left, iliac vein right, small bowel, duodenum, colon, rib left 1, rib left 2, rib left 3, rib left 4, rib left 5, rib left 6, rib left 7, rib left 8, rib left 9, rib left 10, rib left 11, rib left 12, rib right 1, rib right 2, rib right 3, rib right 4, rib right 5, rib right 6, rib right 7, rib right 8, rib right 9, rib right 10, rib right 11, rib right 12, humerus left, humerus right, scapula left, scapula right, clavicula left, clavicula right, femur left, femur right, hip left, hip right, sacrum, face, gluteus maximus left, gluteus maximus right, gluteus medius left, gluteus medius right, gluteus minimus left, gluteus minimus right, autochthon left, autochthon right, iliopsoas left, iliopsoas right, urinary bladder

### S2: List of all pretrained models which were used during data annotation

**Name**: Multi-Atlas Labeling Beyond the Cranial Vault - Abdomen (nnU-Net Task 17) (https://doi.org/10.7303/syn3193805)
**Classes**: spleen, kidney left/right, gallbladder, esophagus, liver, stomach, aorta, inferior vena cava, portal vein and splenic vein, pancreas, adrenal gland left/right

**Name**: RibFrac 2020 Challenge (https://ribfrac.grand-challenge.org/dataset, https://doi.org/10.1016/j.ebiom.2020.103106)

**Classes**: 24 ribs (the dataset only provides segmentations of all ribs; with postprocessing we split this into instance segmentations for each rib).

**Name**: Large Scale Vertebrae Segmentation Challenge (https://github.com/anjany/verse, https://doi.org/10.1016/j.media.2021.102166)

**Classes**: 25 vertebrae

**Name**: Johof Lung Segmentation (https://github.com/JoHof/lungmask, https://doi.org/10.1186/s41747-020-00173-2)

**Classes**: 5 lung lobes

**Name**: SegTHOR: Segmentation of THoracic Organs at Risk in CT images (nnU-Net Task 55) (https://competitions.codalab.org/competitions/21145, https://ceur-ws.org/Vol-2349/)

**Classes**: aorta, esophagus, heart, trachea

**S3: More details on data annotation**

*Transfer of segmentations between CTs with and without contrast agent*

For segmentation of the different heart subparts (left/right atrium, left/right ventricle, myocardium, pulmonary artery) on all different CT series (also without contrast agent) the following approach was taken: An inhouse dataset was available with ground truth segmentations of the heart subparts on CT images with arterial contrast agent application. For each image with contrast agent, a well aligned native CT image was also available. The segmentation was transferred to the native CT images and a nnU-Net was trained on images with and without contrast agent. This model was then used to predict the heart subparts in our dataset, resulting in segmentations of the heart subparts in all kinds of CT sequences.

**S4: Model details**

Two adaptations to the default nnU-Net settings were applied: First, mirroring was removed from the data augmentation pipeline because otherwise, the model was not able to properly distinguish between left and right anatomical structures. Second, the number of training epochs was increased from the default of 1000 to 4000 given the large size of the dataset.

Training one nnU-Net model for all 104 classes at the same time results in very high memory consumption. Thus, we split the 104 classes into five parts, four with 21 classes and one with 20 classes. For each part, one smaller nnU-Net was trained.

For the 3 mm model it was possible to combine all 104 classes into one model. Training epochs were increased even further to 8000 since training took longer given the high number of classes in one model for the low-resolution model).

**S5: List of 12 structures part of the "Multi-Atlas Labeling Beyond the Cranial Vault Challenge" which we use as baseline**

- Aorta
- Gallbladder
- Inferior vena cava
- Left adrenal gland
- Left kidney
- Liver
- Pancreas

- Portal vein and splenic vein

- Right adrenal gland

- Right kidney

- Spleen

- Stomach

**S6: Reasons for low performance on some structures**

Overall, the proposed model shows highly accurate and robust results for most structures. For some structures, however, minor errors could be identified frequently (figure 4), e.g.:

- neighboring ribs or vertebrae are labelled incorrectly: Since the total number of ribs and vertebrae can vary, physicians typically resort to counting the number of ribs/vertebrae from top to bottom to identify individual ribs/vertebrae. However, on some images only a subset of ribs/vertebrae is visible making it difficult for physicians as well as for the model to identify the correct rib numbers or vertebrae number.

- iliac vessel segmentation is missing parts: The dataset also contains images without contrast agent. On those images, the iliac vessels can be hardly visible to the human eye which is also reflected in the model's performance.

- colon segmentation is missing parts: The colon's shape, size, texture and position in the abdomen can vary greatly and is sometimes difficult to distinguish from the small bowel.

- heart subpart segmentation is inaccurate for native images: Since the different heart chambers are difficult to see on native images, the segmentation can be in inaccurate in these cases.

## S7: Details of evaluation on BTCV dataset

For better comparison we resampled the BTCV dataset, which contains 30 subjects, to the same 1.5 mm isotropic resolution as our training dataset. Then we trained a nnU-Net with 5-fold crossvalidation on the resampled BTCV dataset. This resulted in predictions for all 30 subjects which we compared our proposed model to. To compare our model to the BTCV model on the test set of our dataset we used the ensemble of the 5 models from the cross-validation to make predictions.

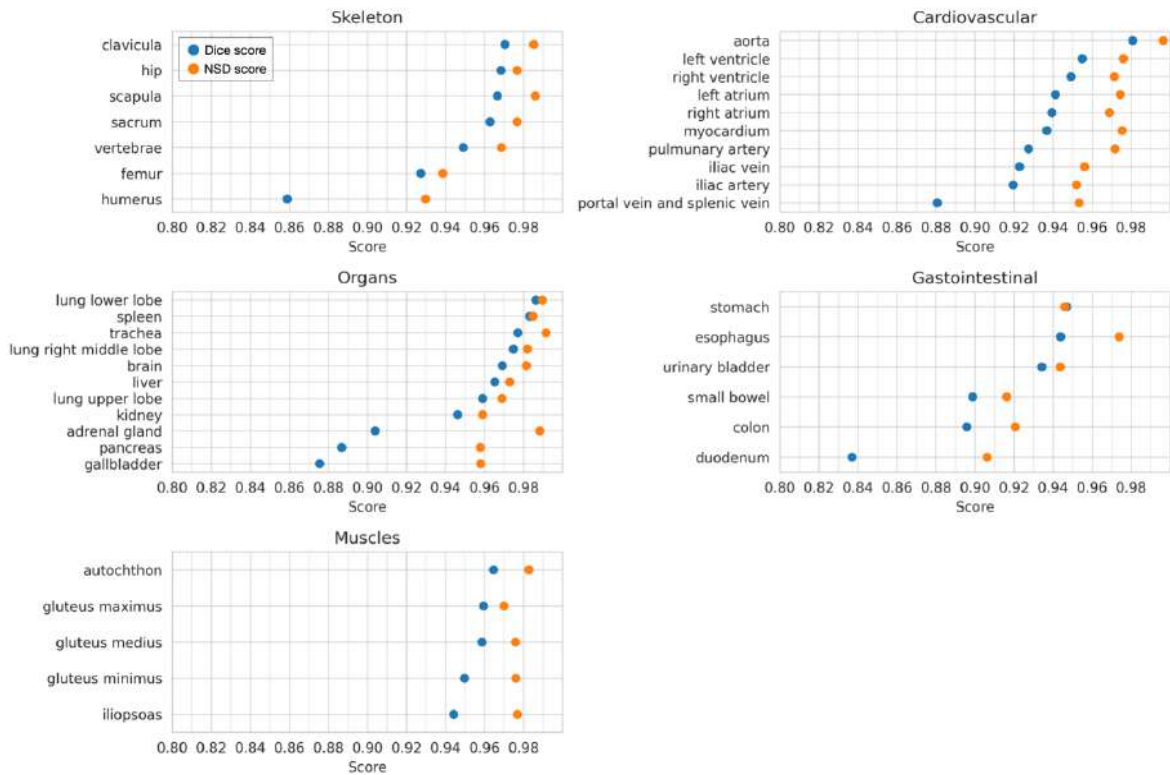

Figure S1: Results of the high-resolution model per class (similar classes, e.g. ribs or left/right of the same class are grouped together) (Blue: Dice score, Orange: NSD score).

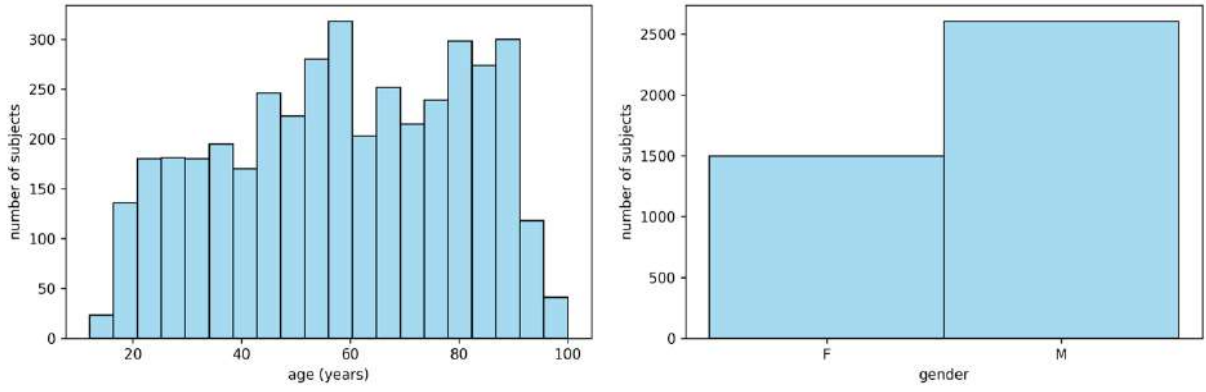

Figure S2: Age distribution of the aging study dataset. The distribution is quite homogeneous since polytraumas occur relatively homogenously over the lifespan.

## Figure S3 – Aging study

Illustration of attenuation in HU values and volume in milliliters (ml) of every of 104 segmented body structures displayed as Spearman ($r_s$) correlation with reported p-values as well as boxplots for each age quartile (Q1: < 41 y; Q2: 41 – 59 y; Q3: 59 – 78 y; Q4: > 78 y). Group comparison was performed using Kruskal-Wallis test and is listed for all body structures in Supplement Table 1. Significant differences between groups determined by post-hoc analysis with Mann-Whitney-U are indicated in the graphs. If the volume of a body structure in a patient was below the lower cut off for each body structure, an error was assumed and the respective body structure of the patient was excluded.

The specific structure is the respective title of the illustration. Concerning the Boxplots the central mark indicates the median, the bottom and top edges of the box indicate the 25th and 75th percentiles, respectively. The whiskers extend to the most extreme data points not considered outliers, which are not displayed to sustain readability.

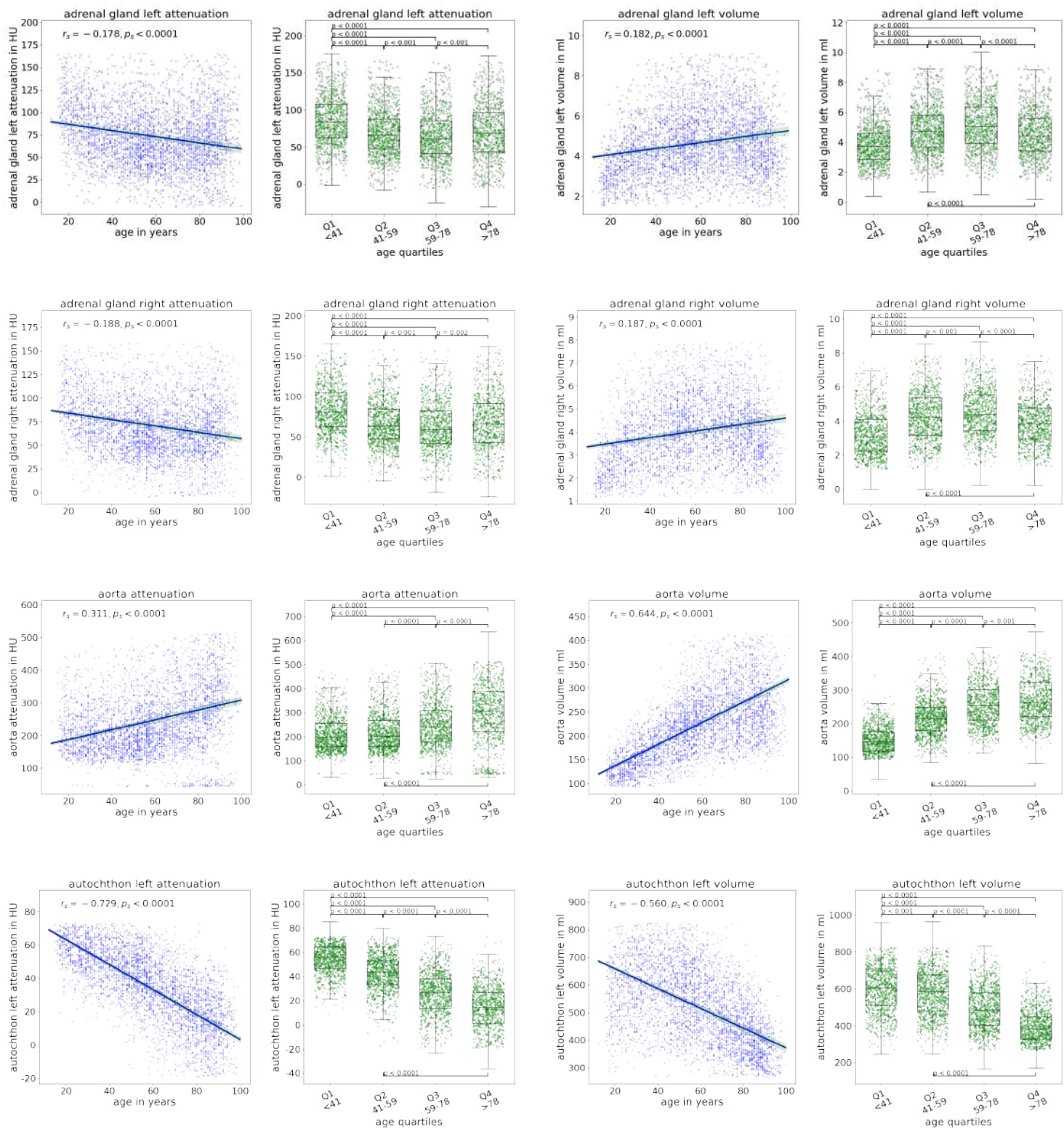

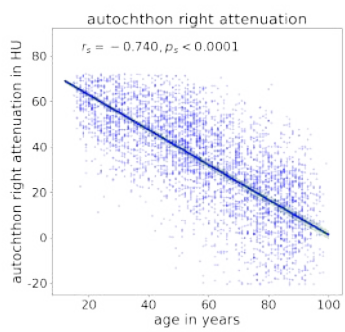
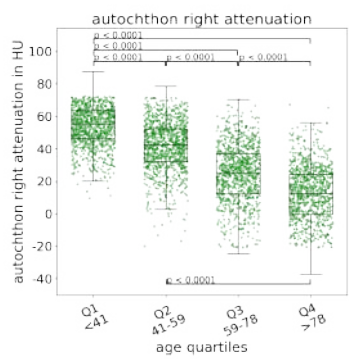
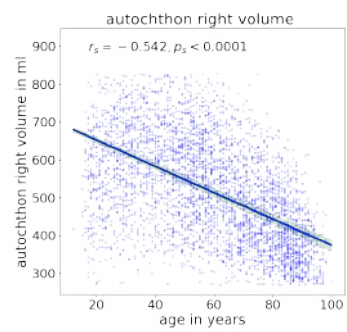
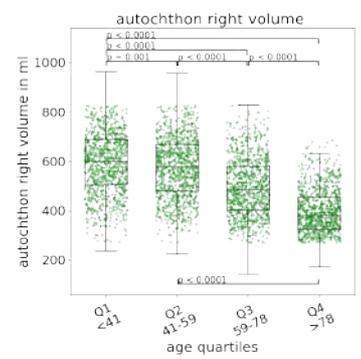
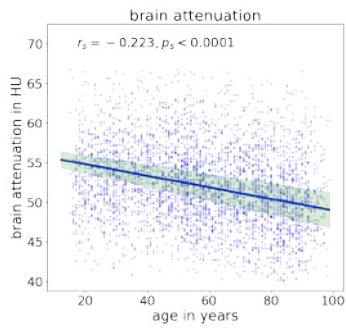
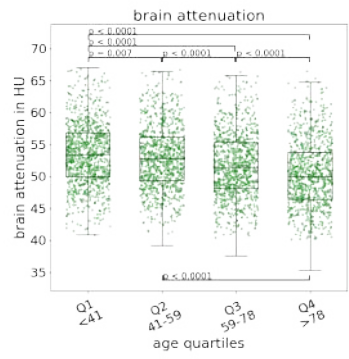
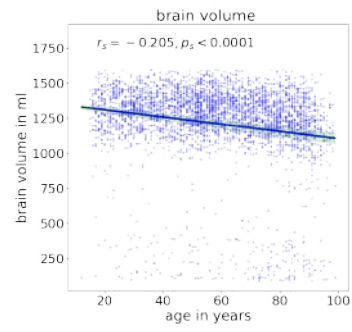
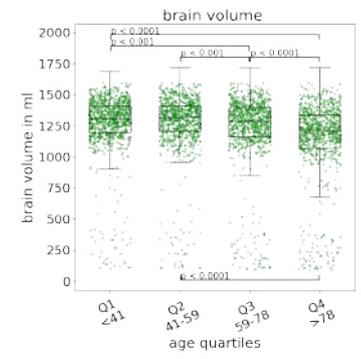
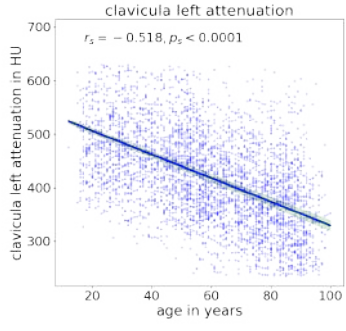
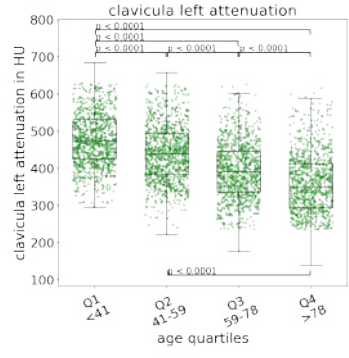
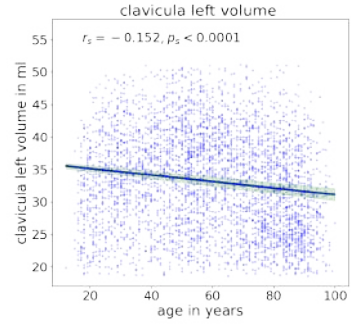
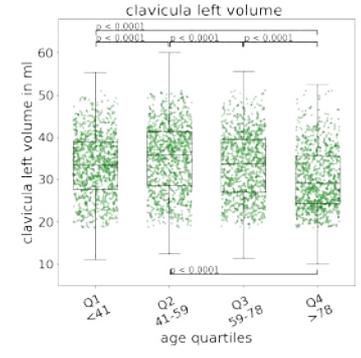
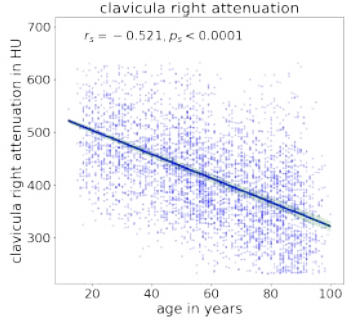
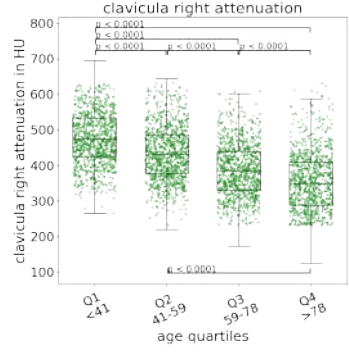
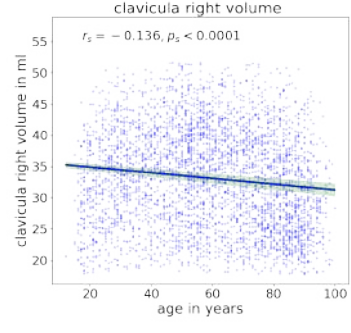
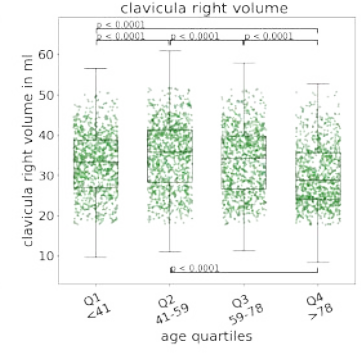
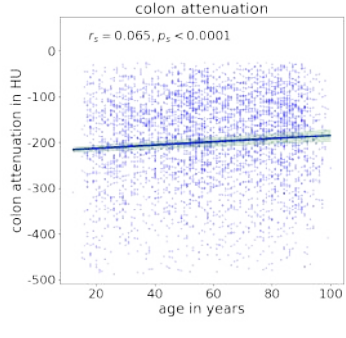
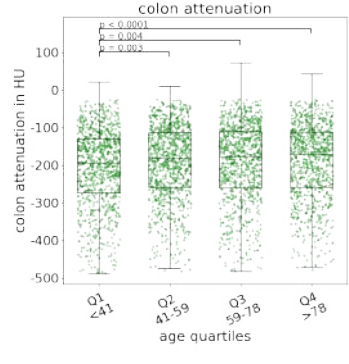
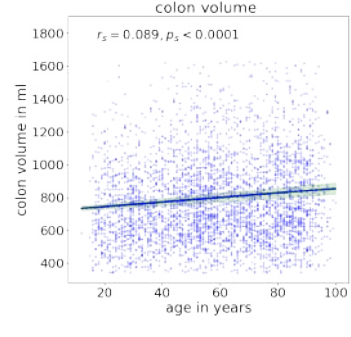
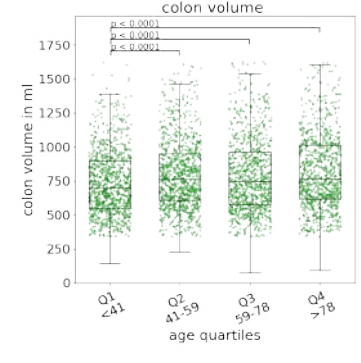

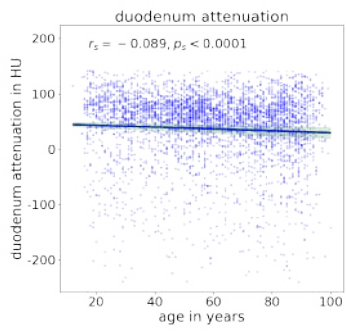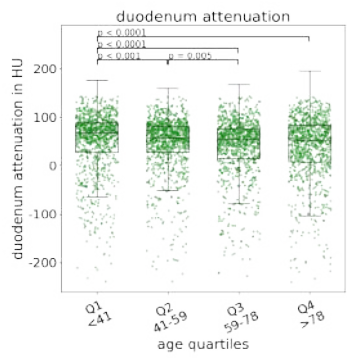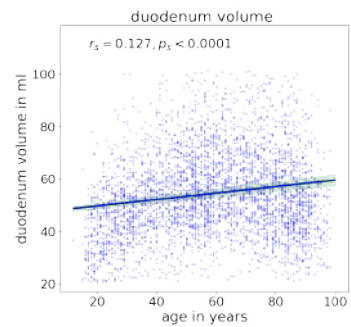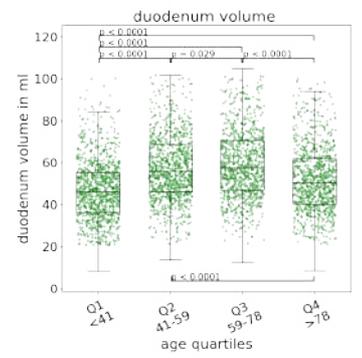
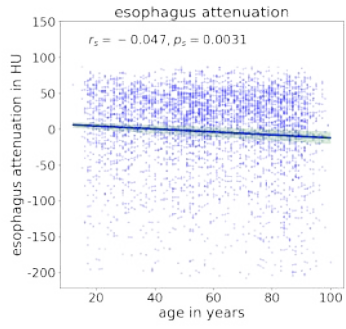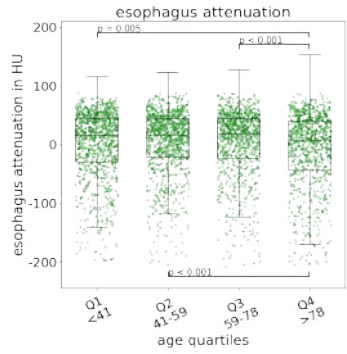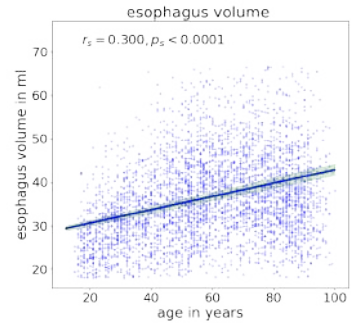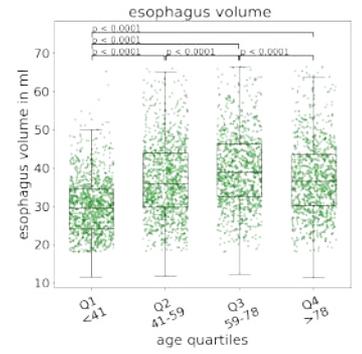
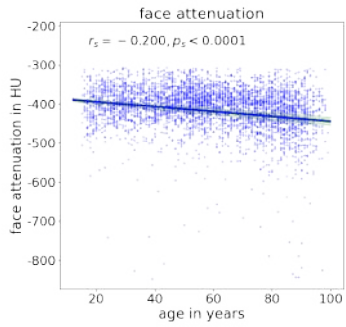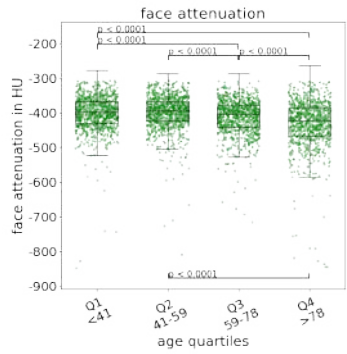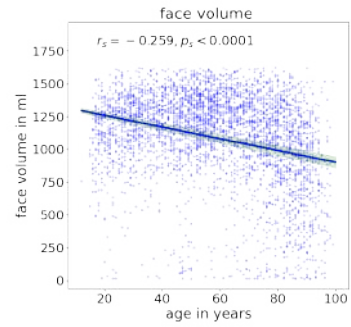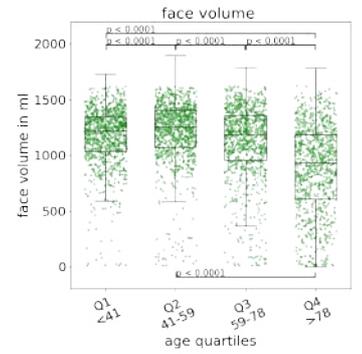
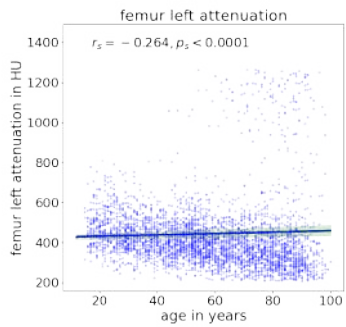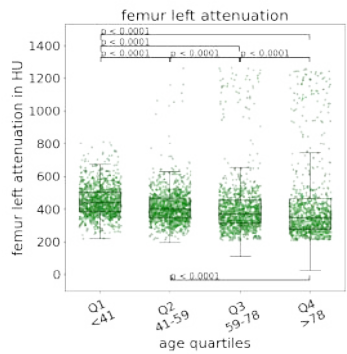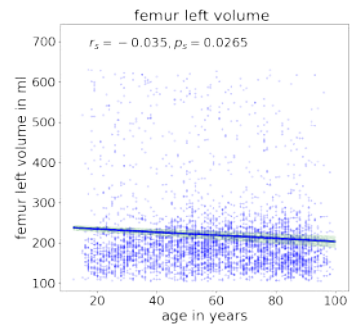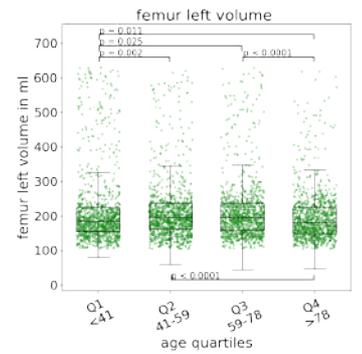
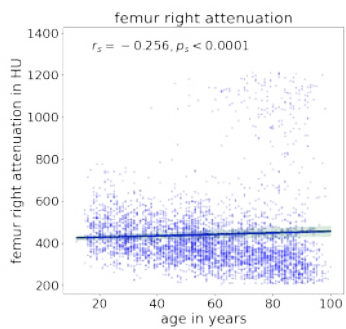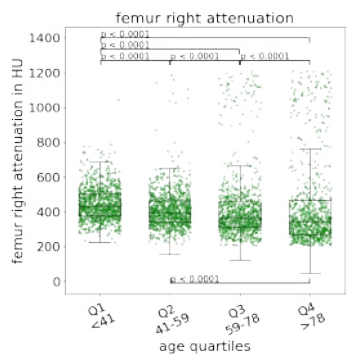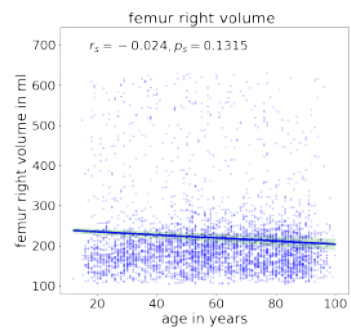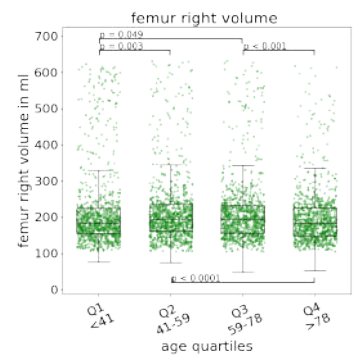

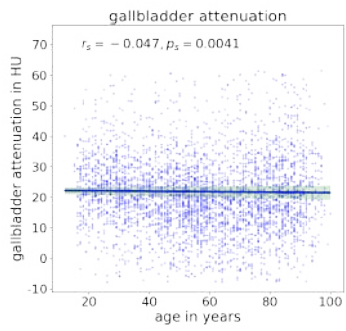 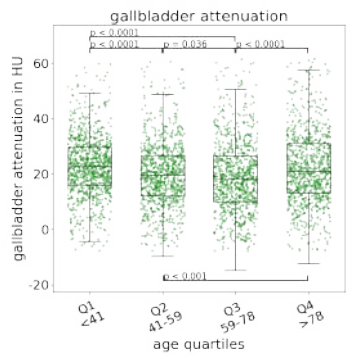 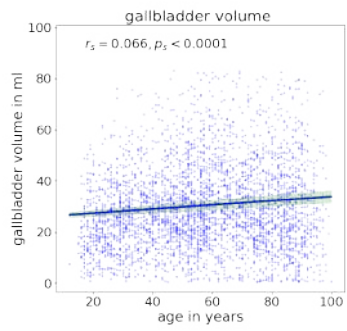 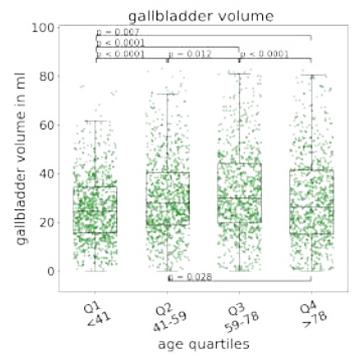
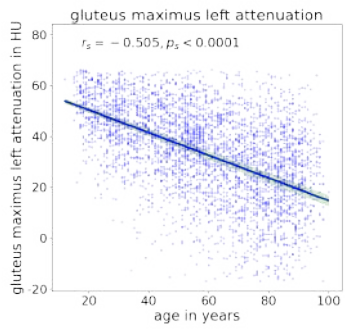 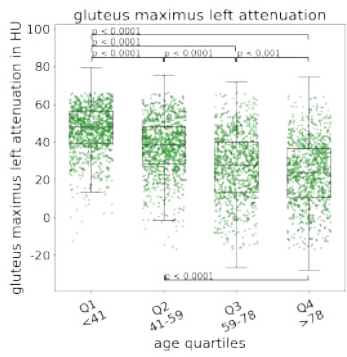 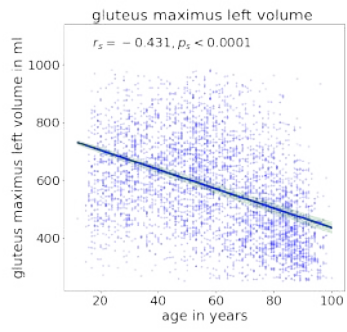 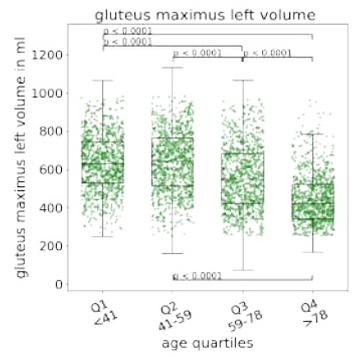
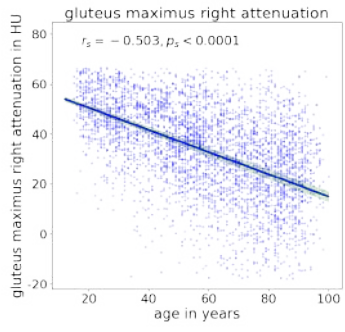 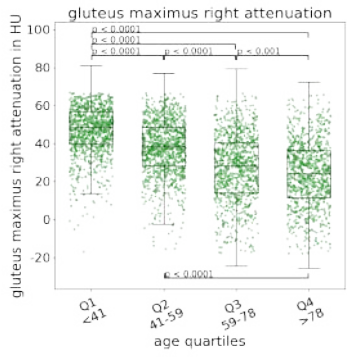 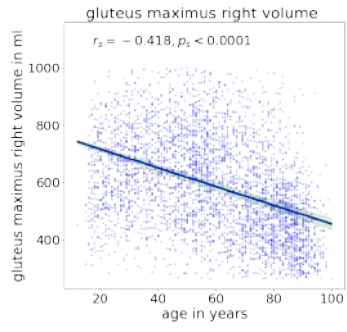 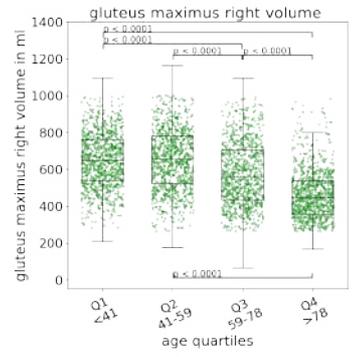
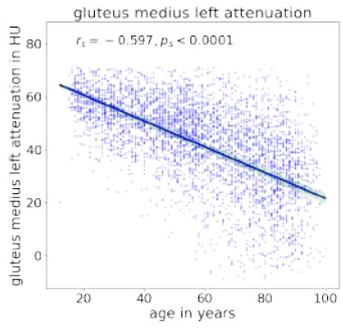 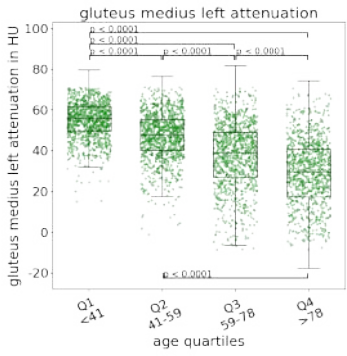 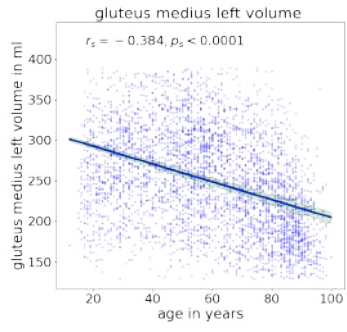 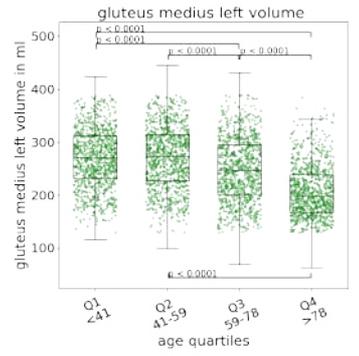
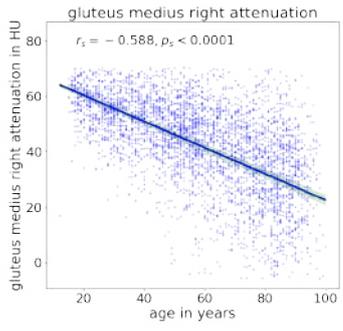 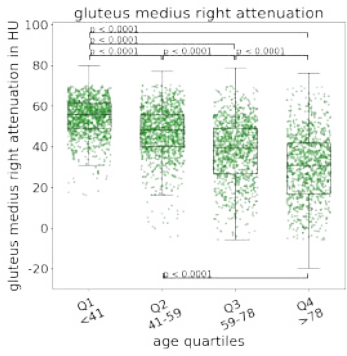 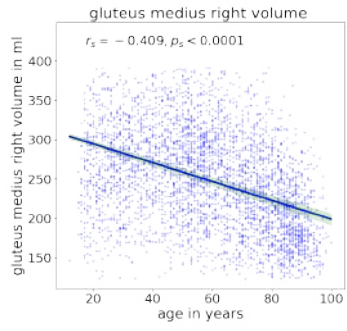 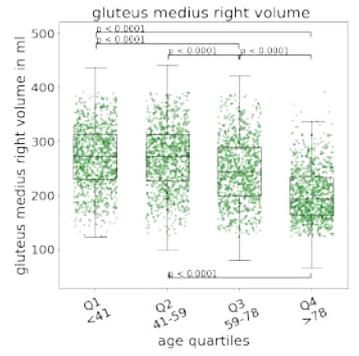

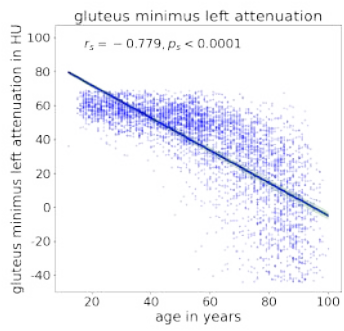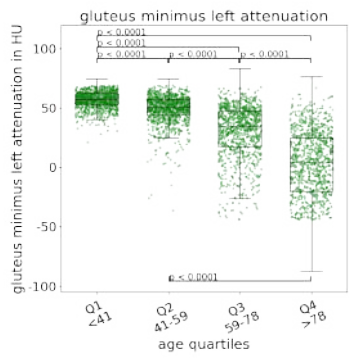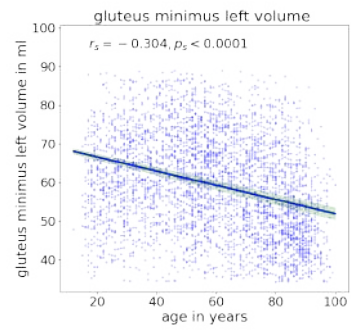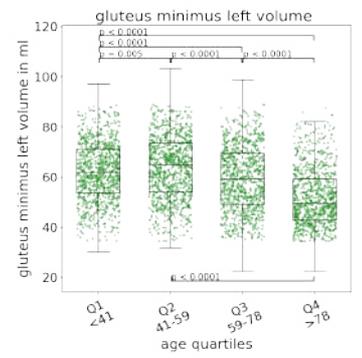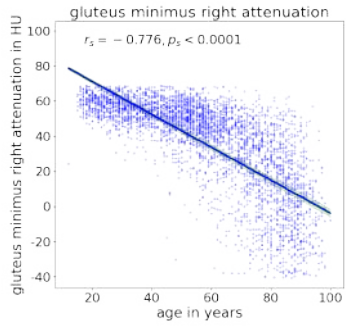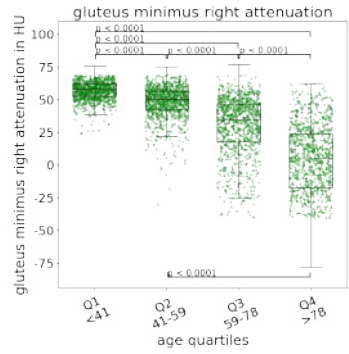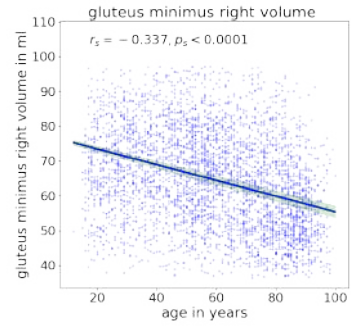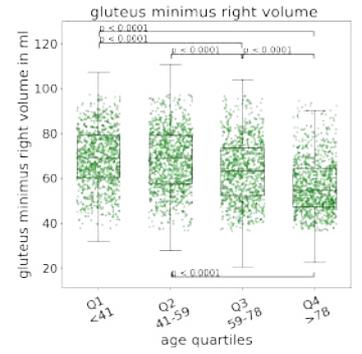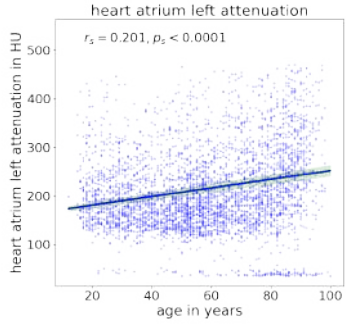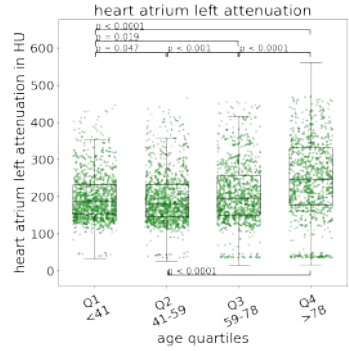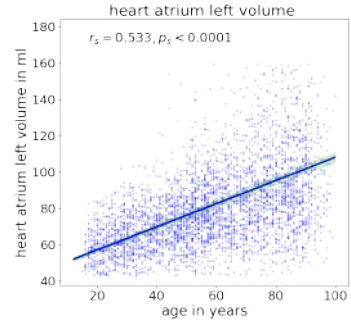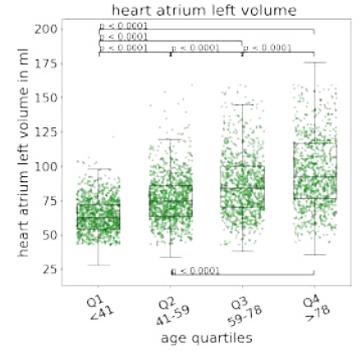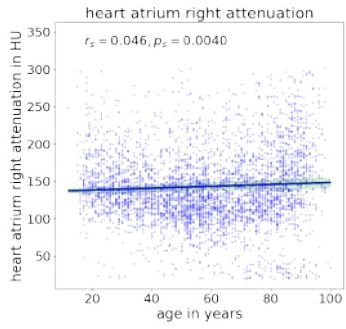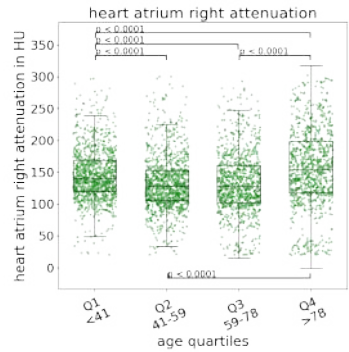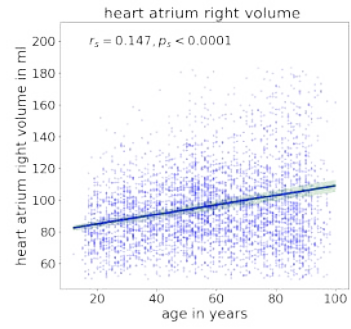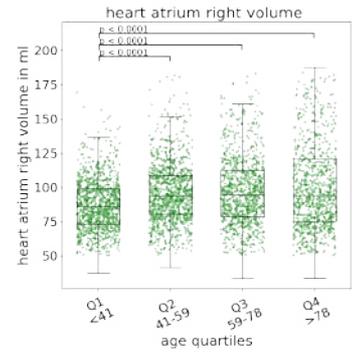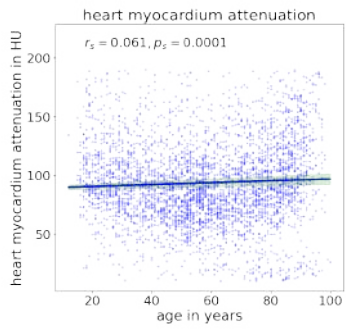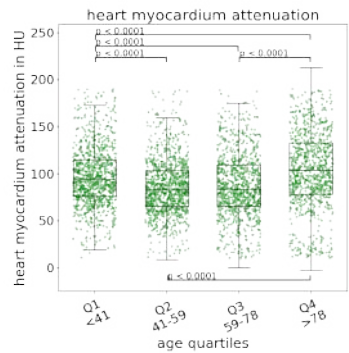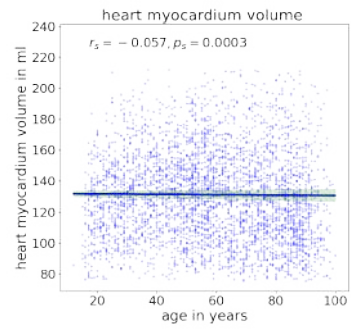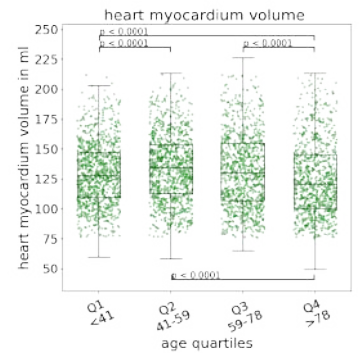

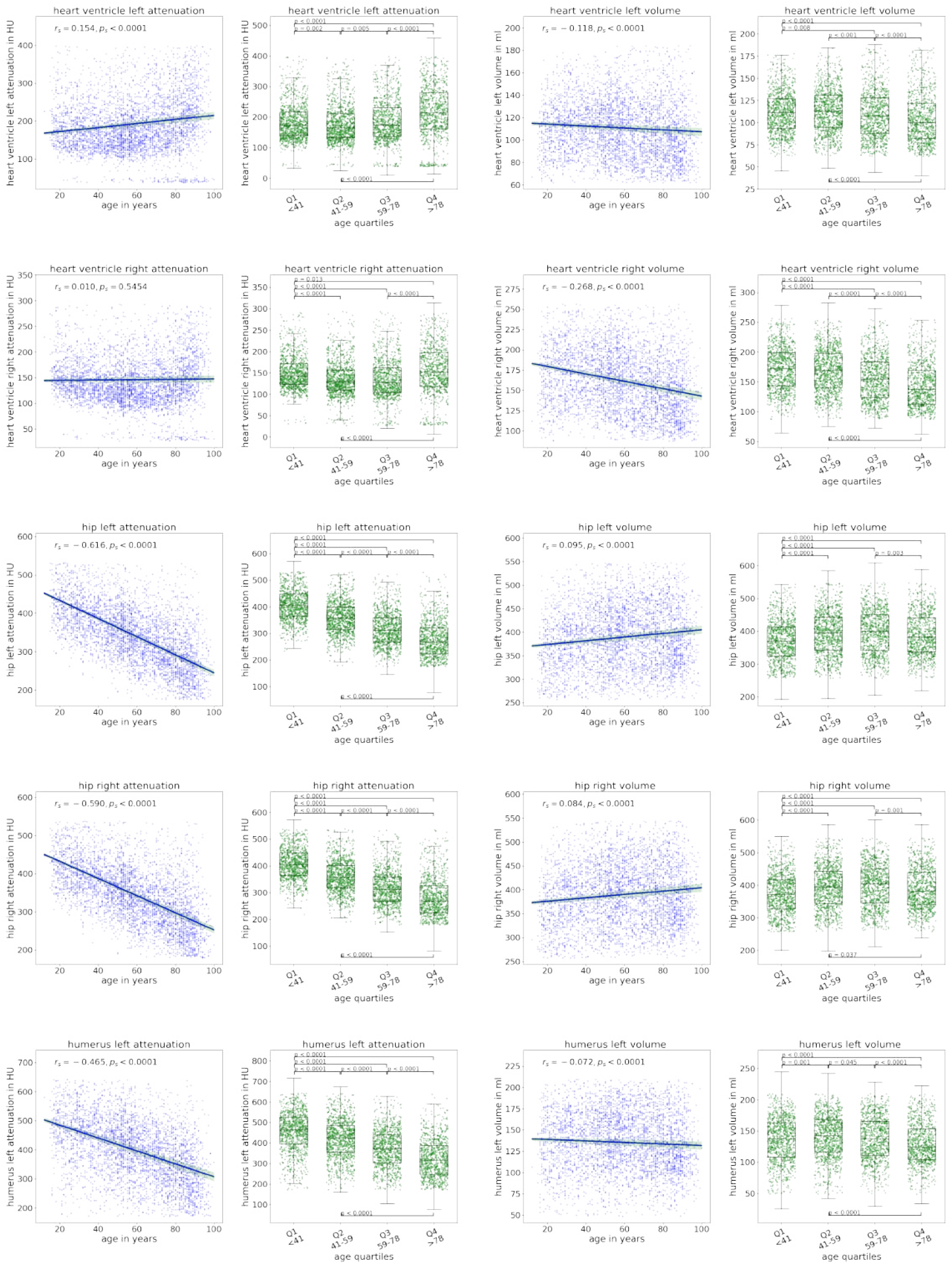

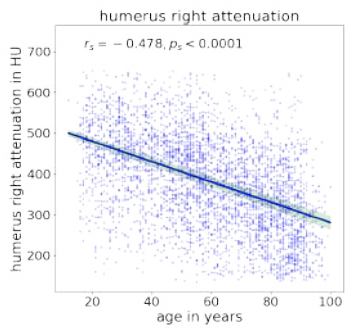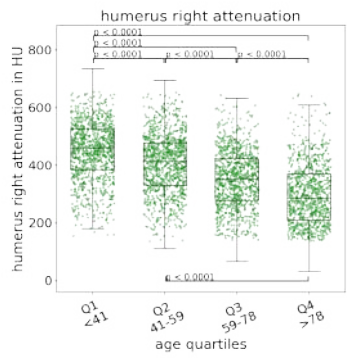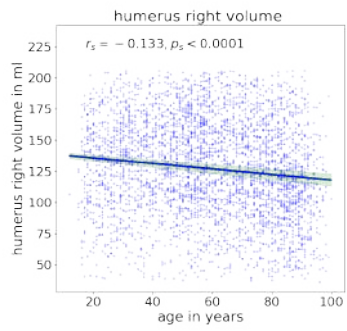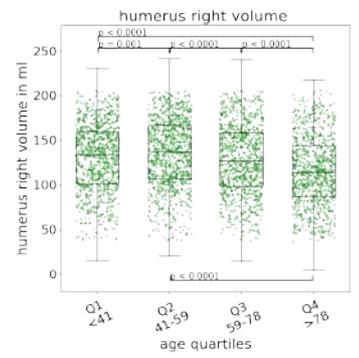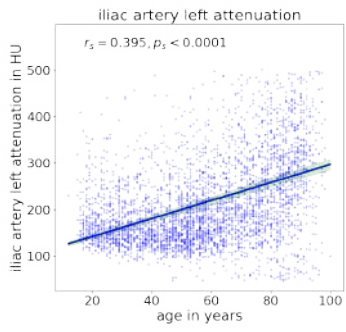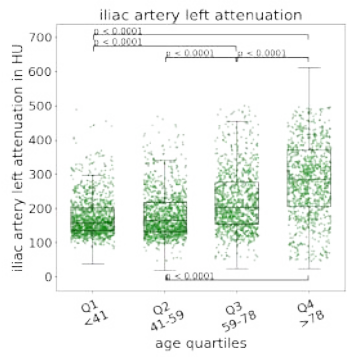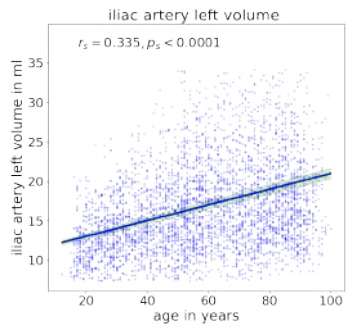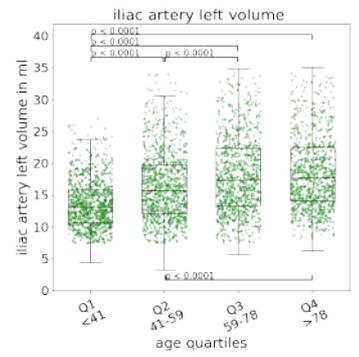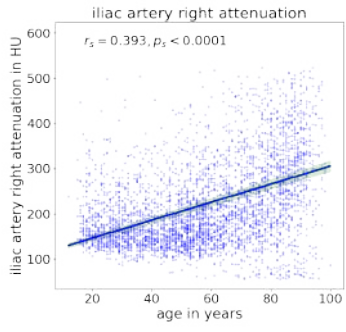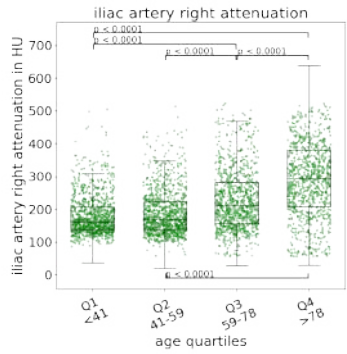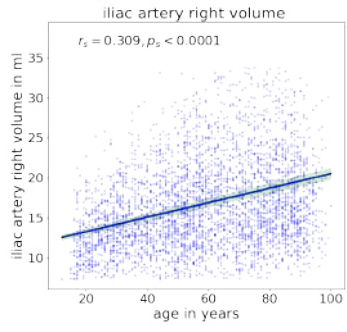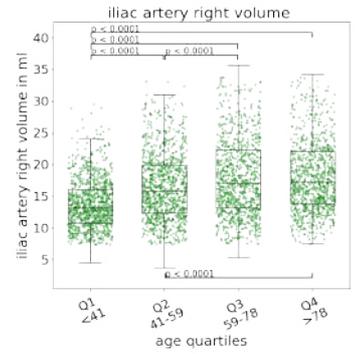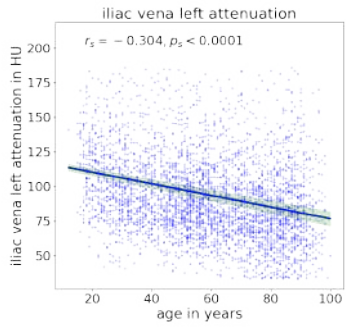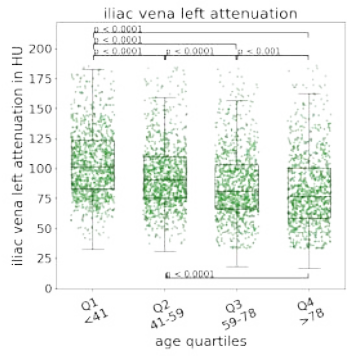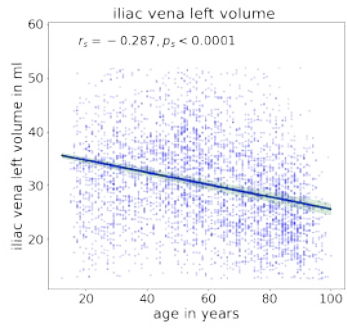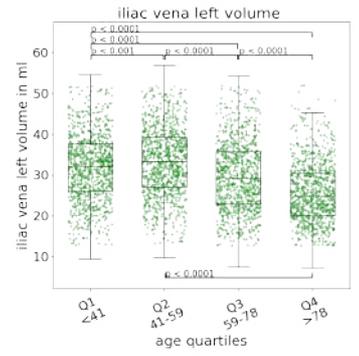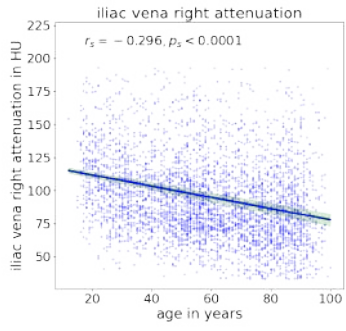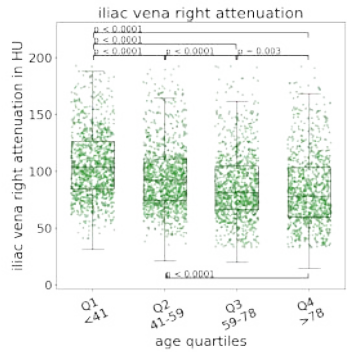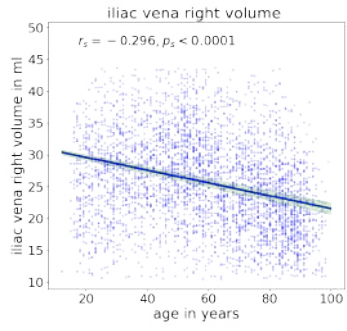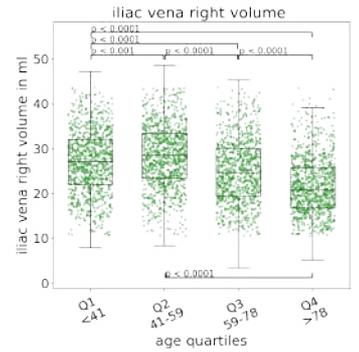

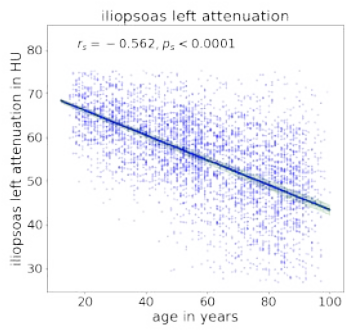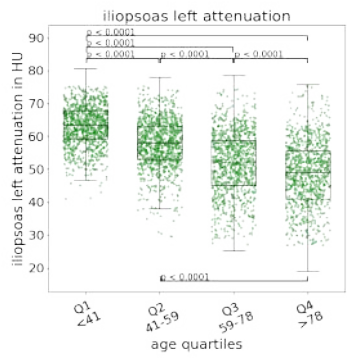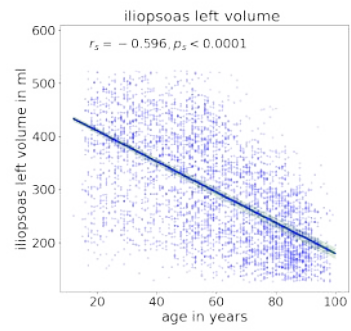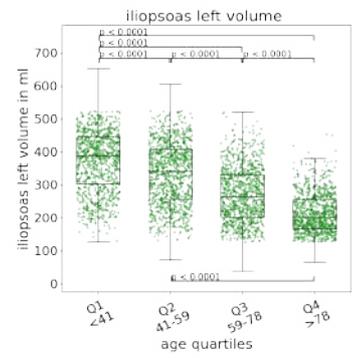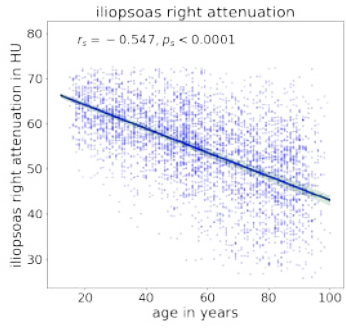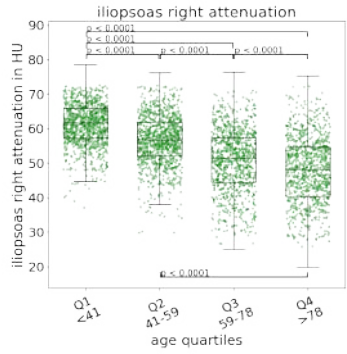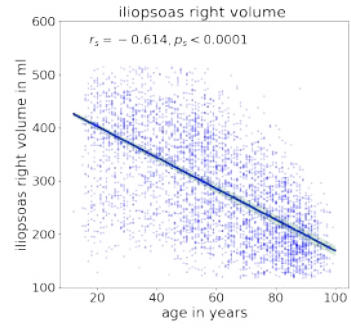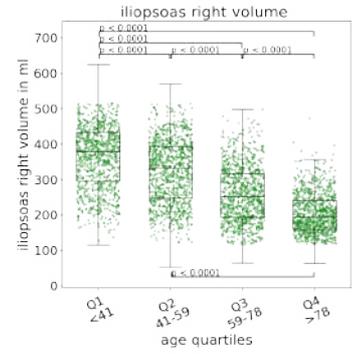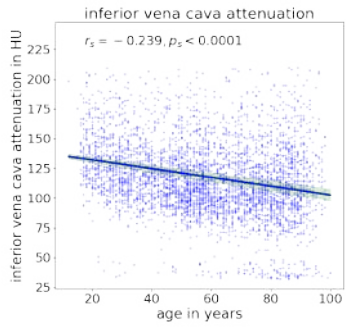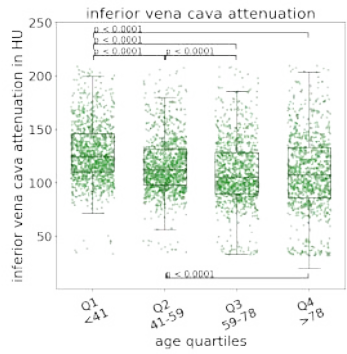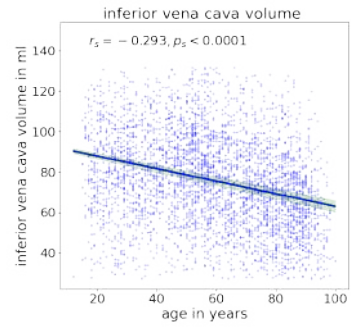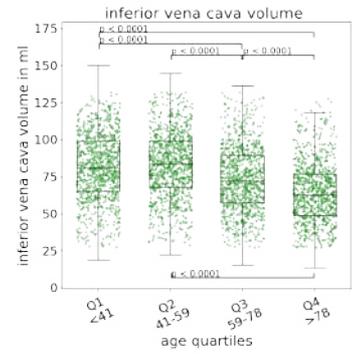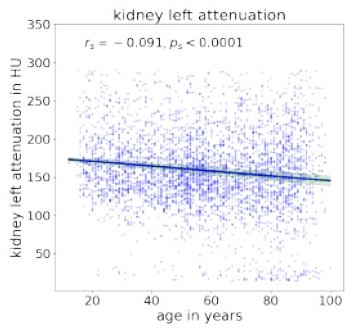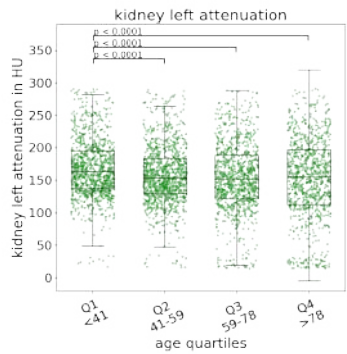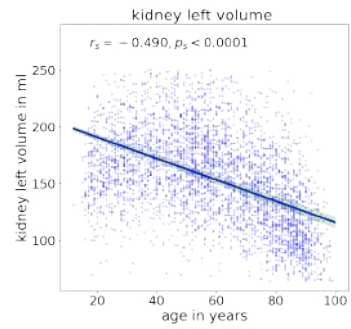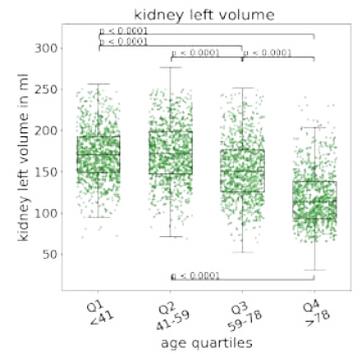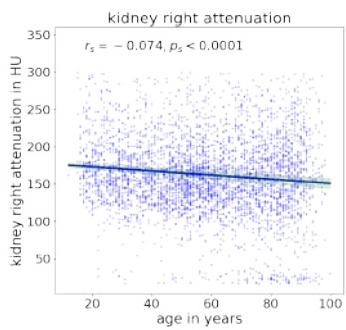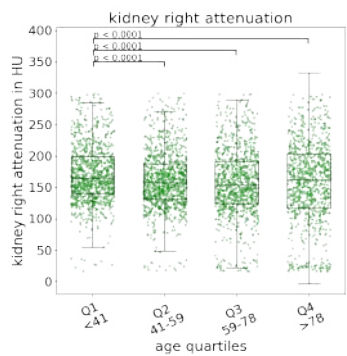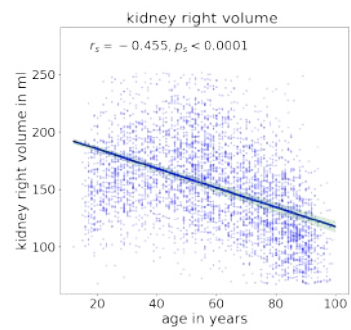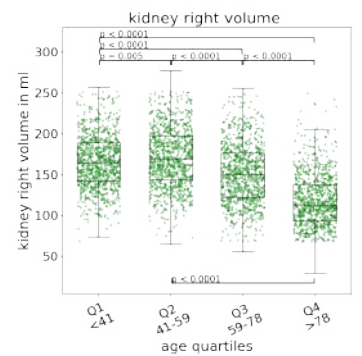

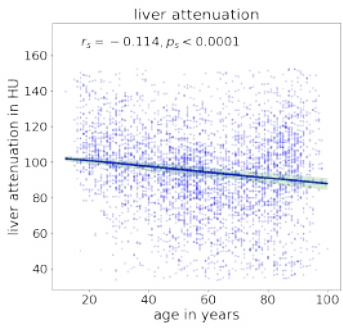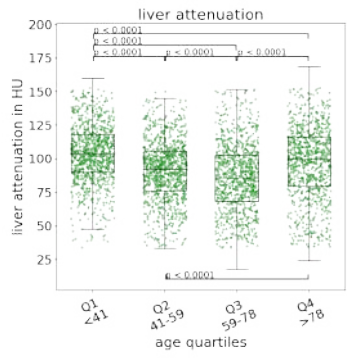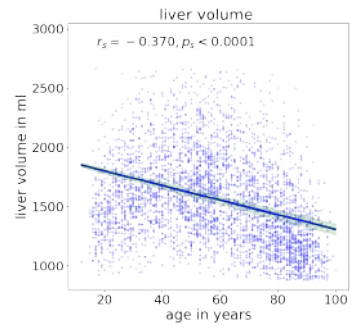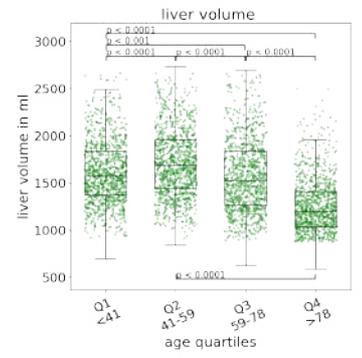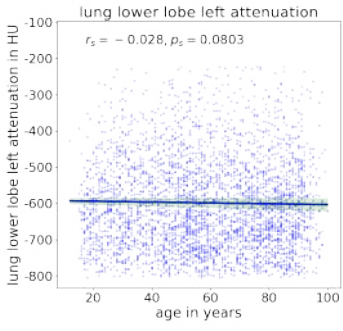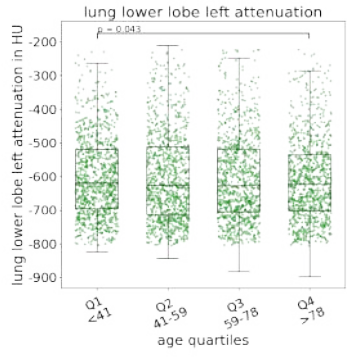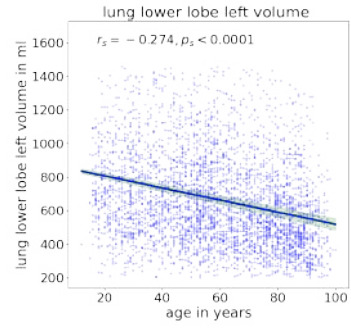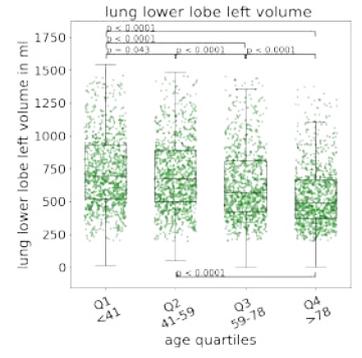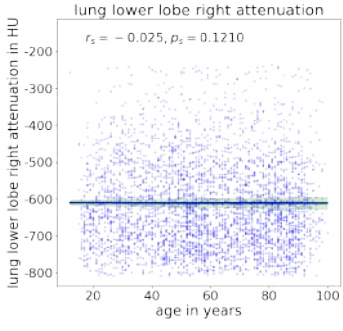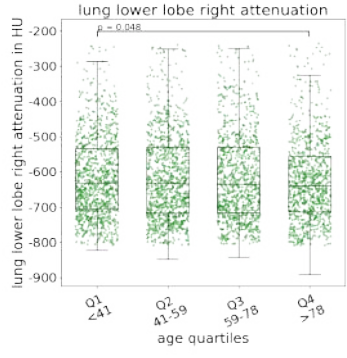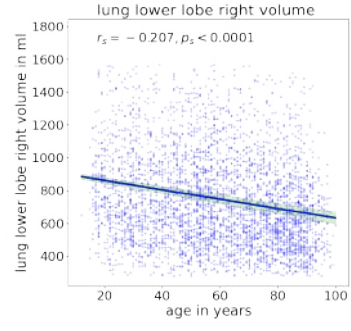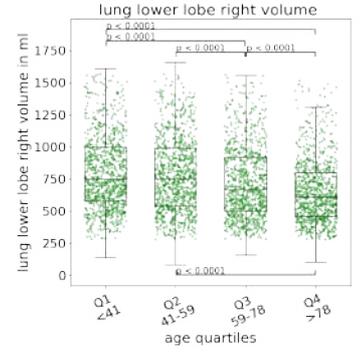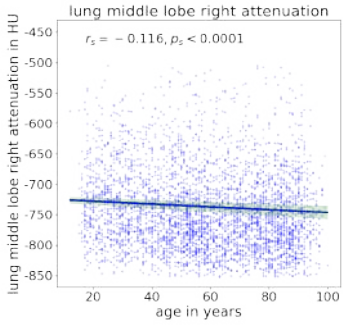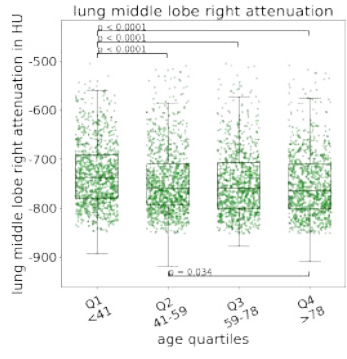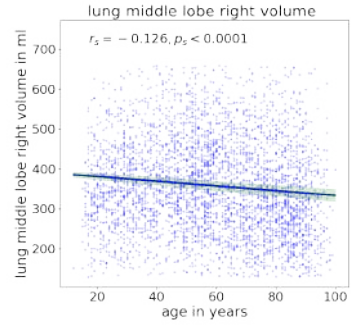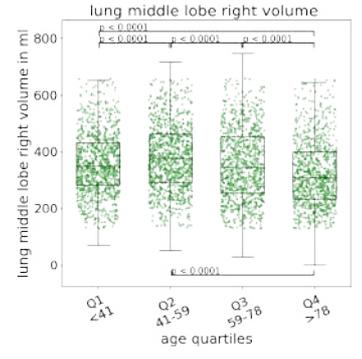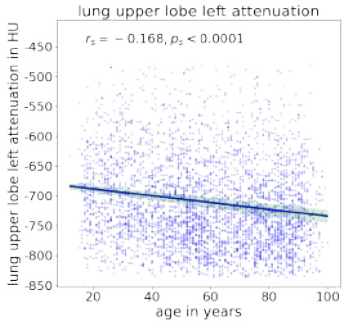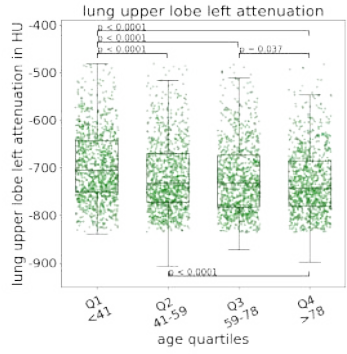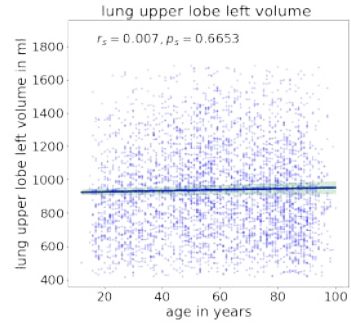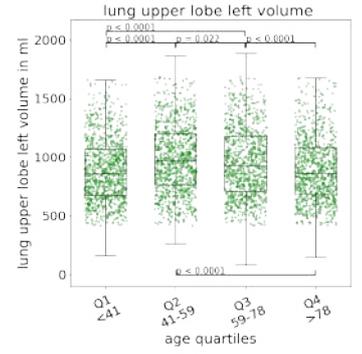

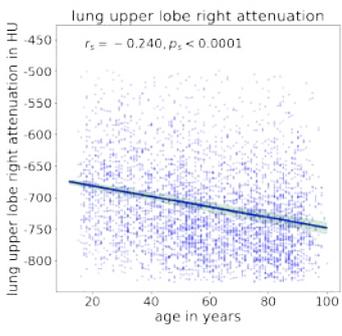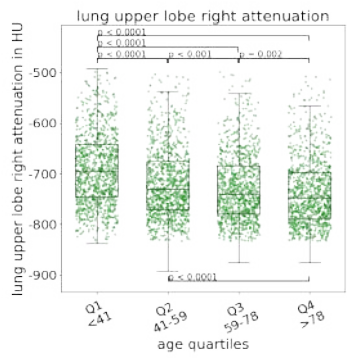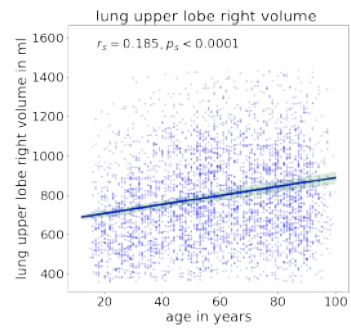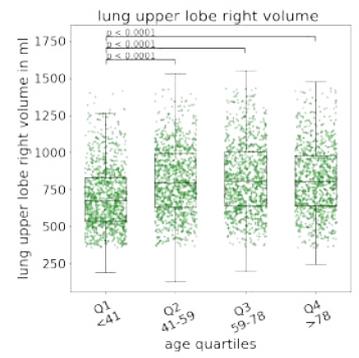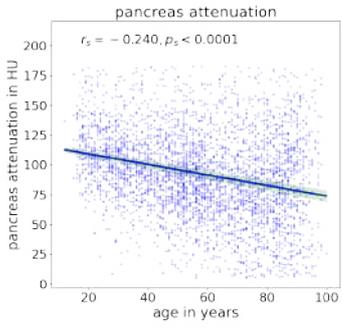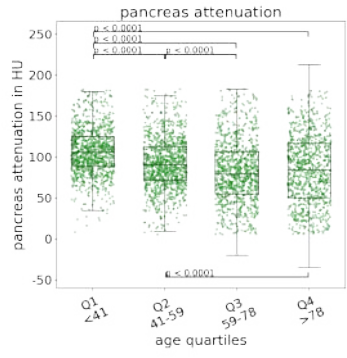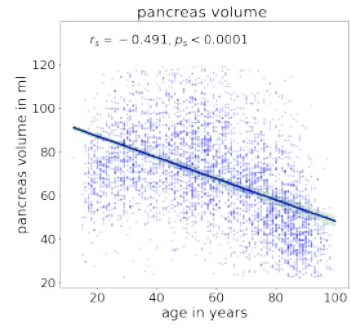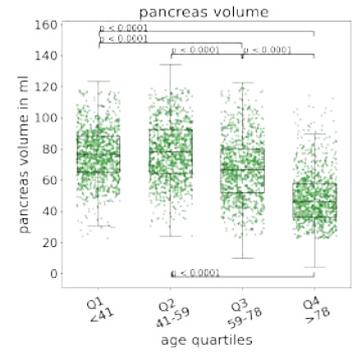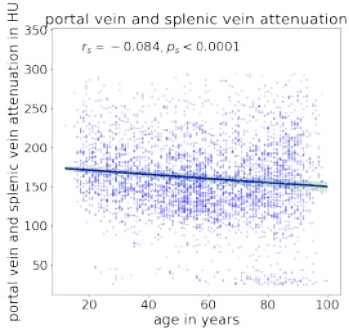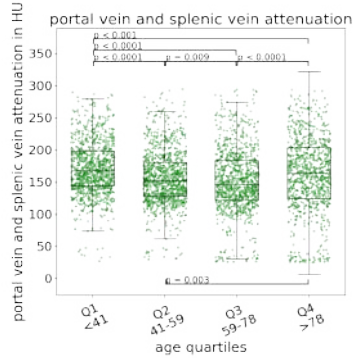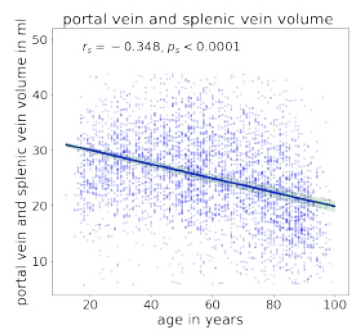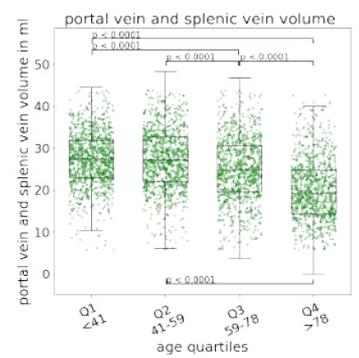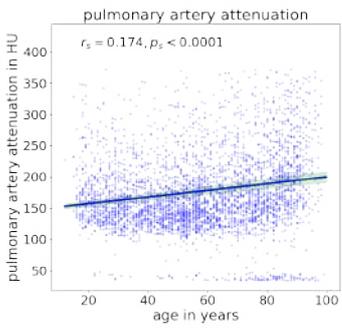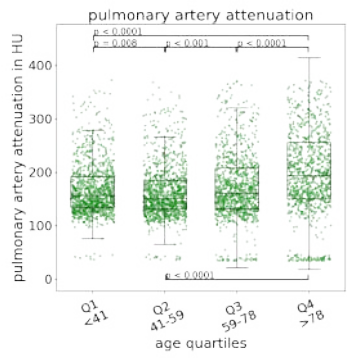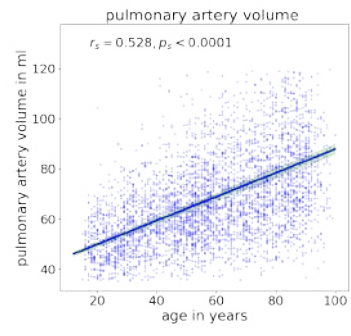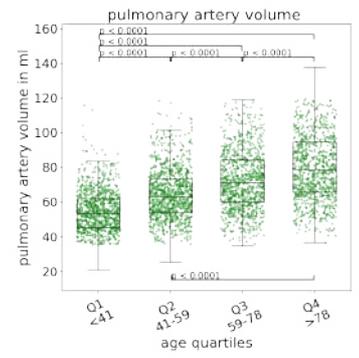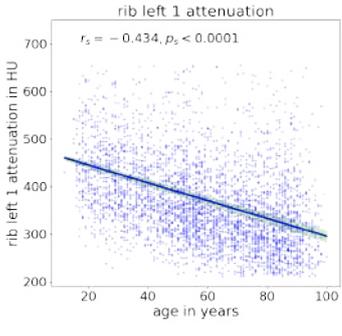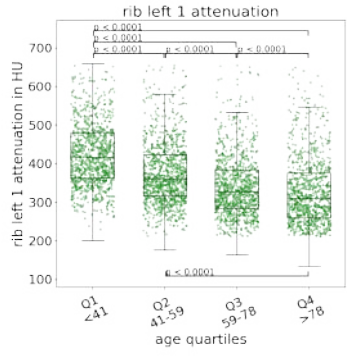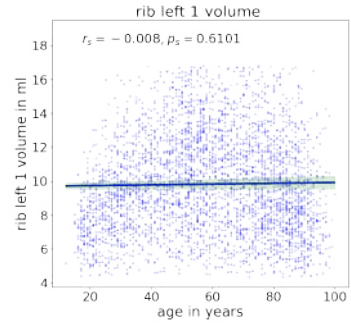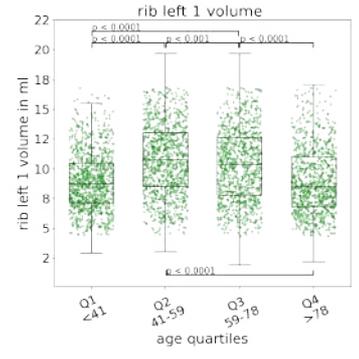

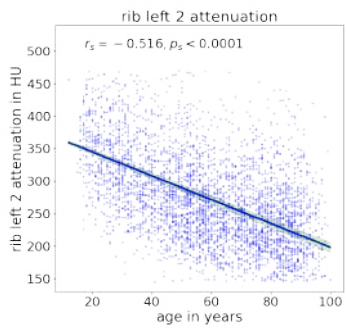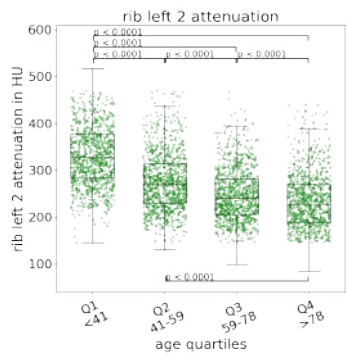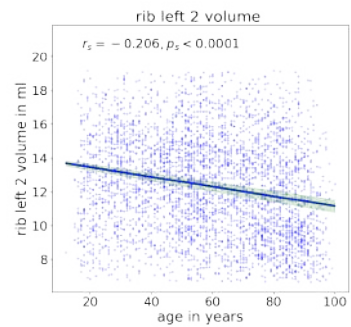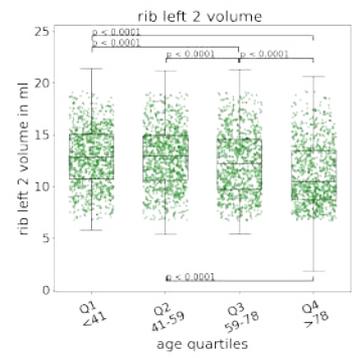
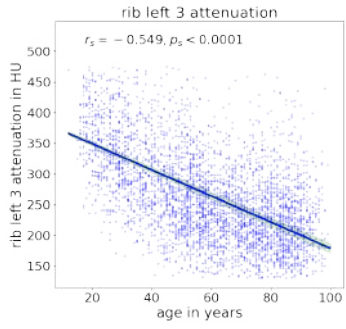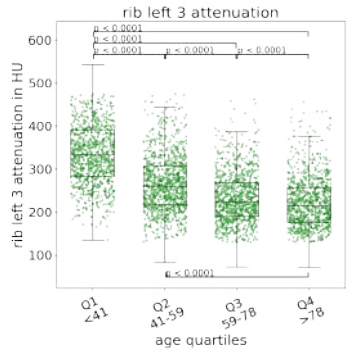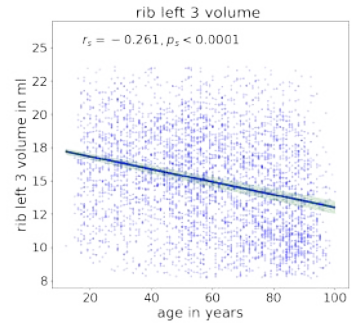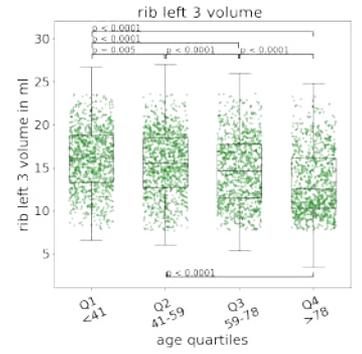
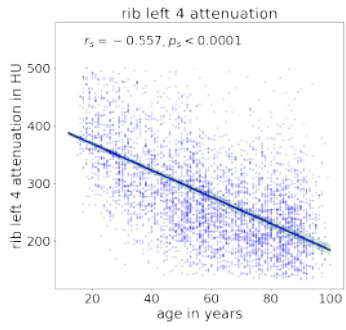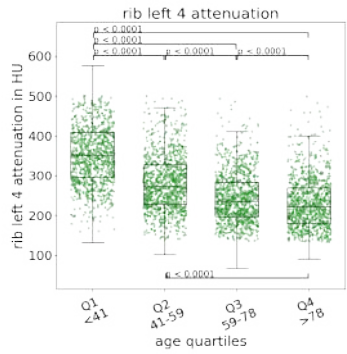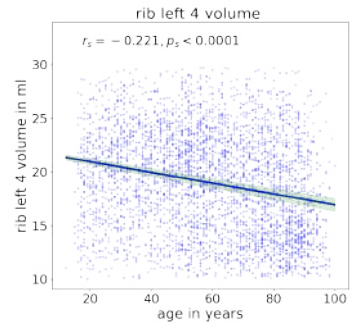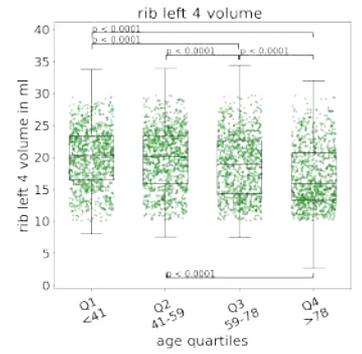
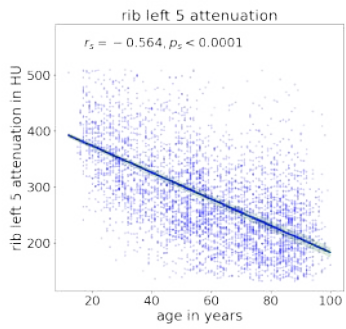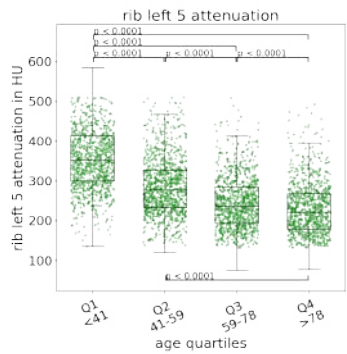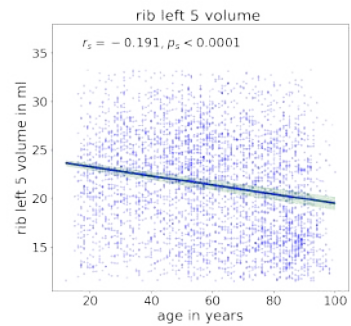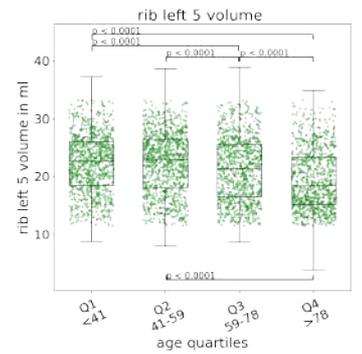
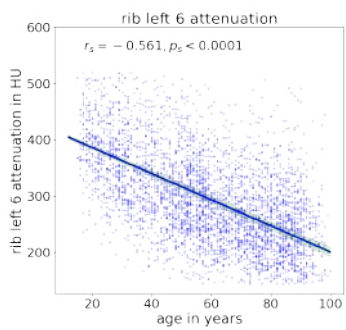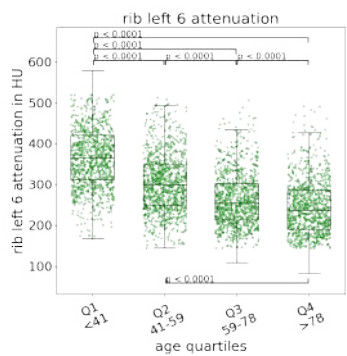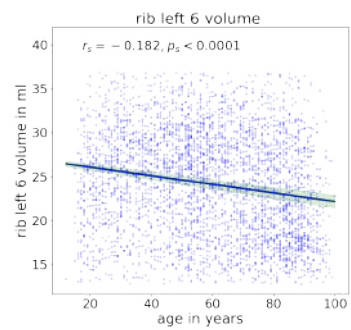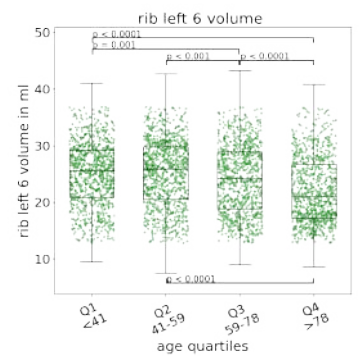

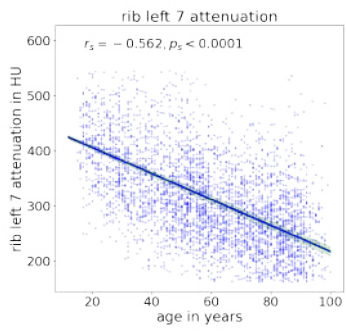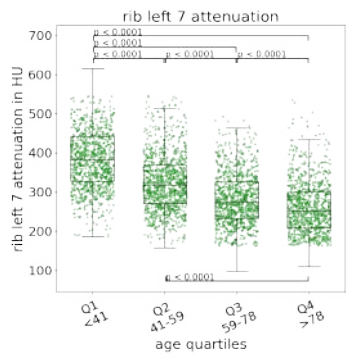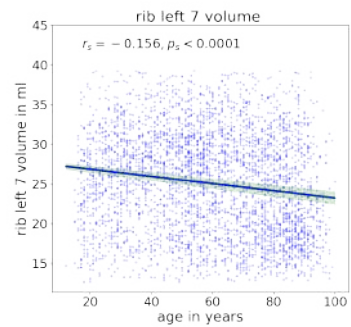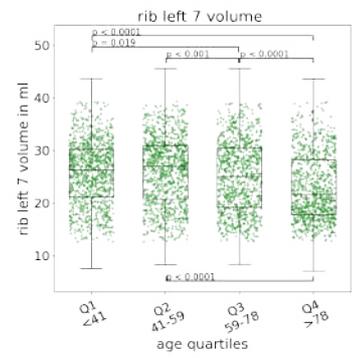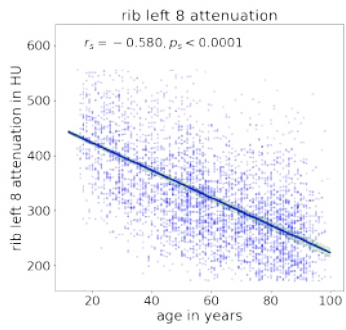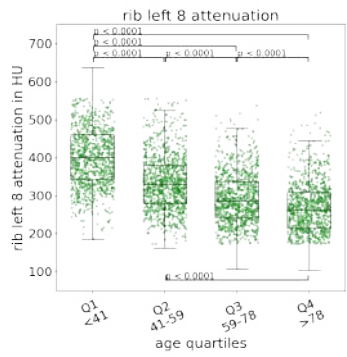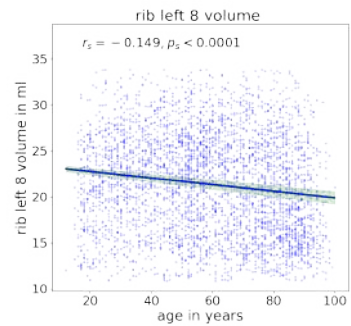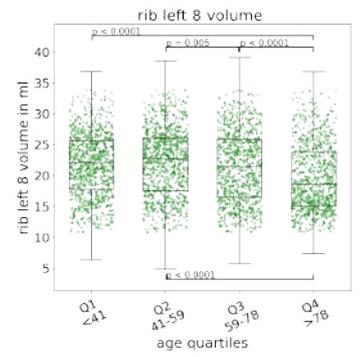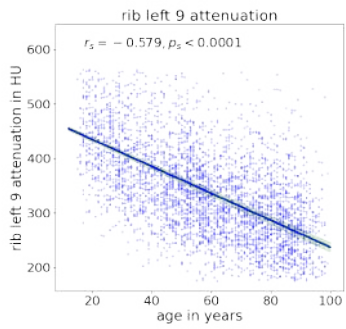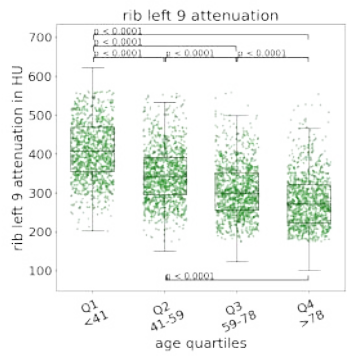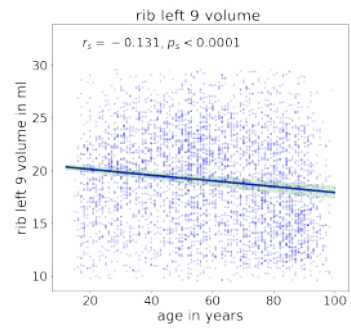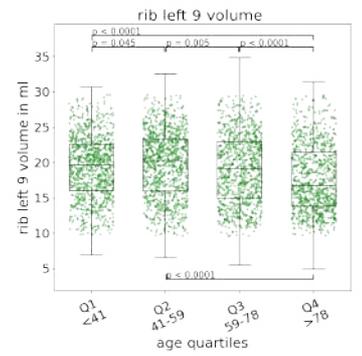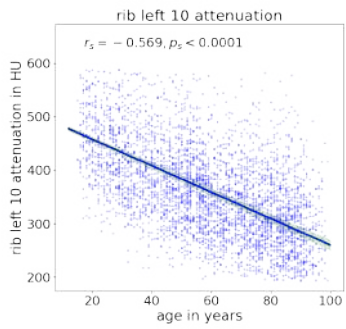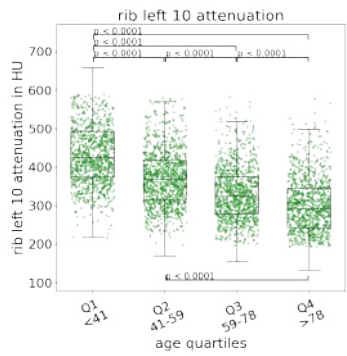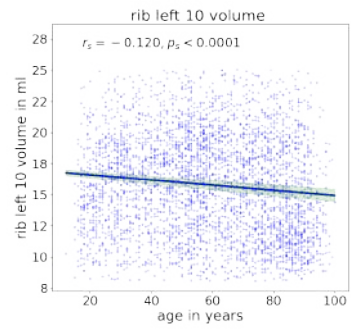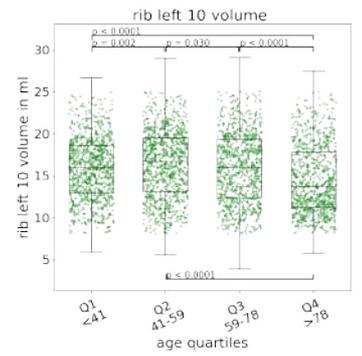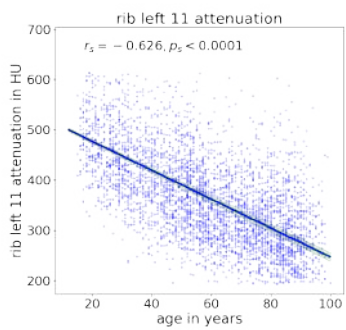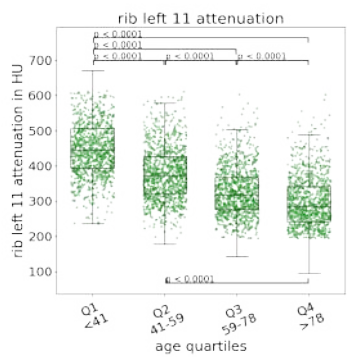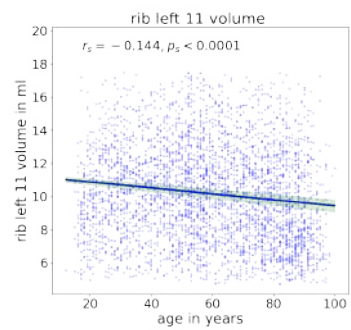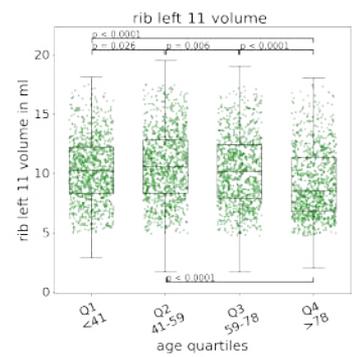

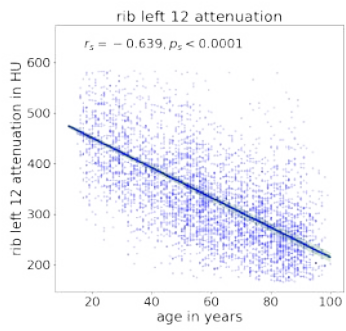 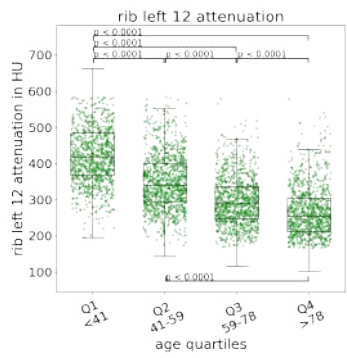 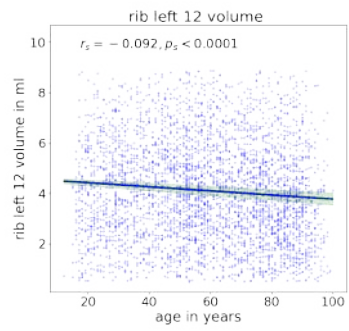 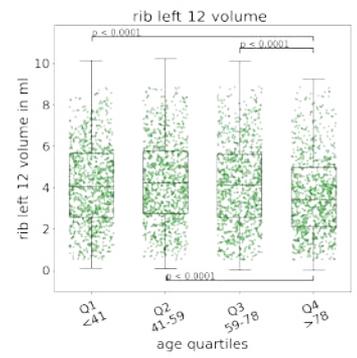
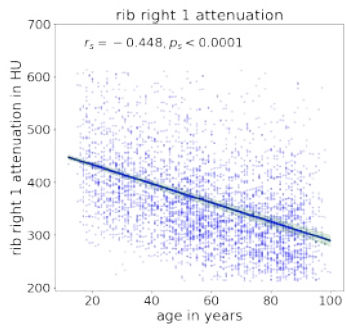 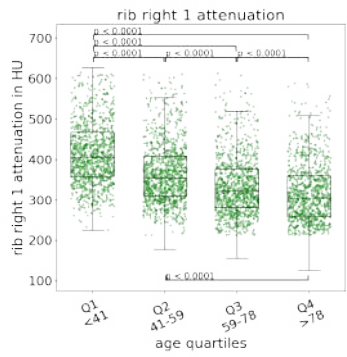 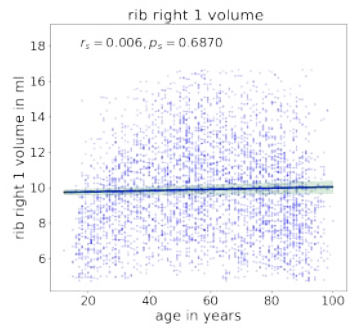 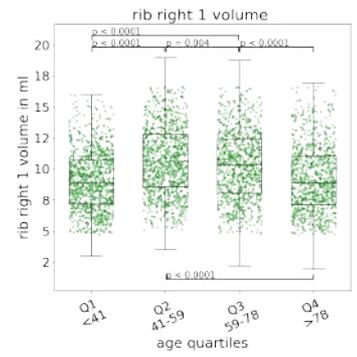
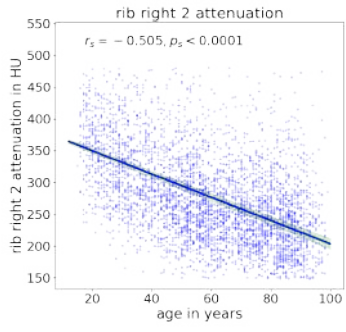 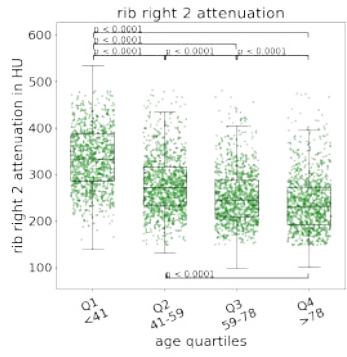 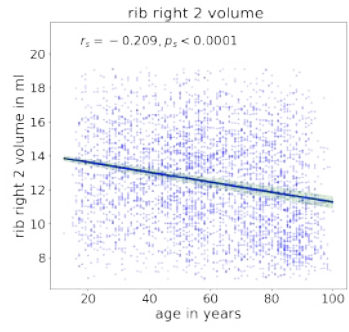 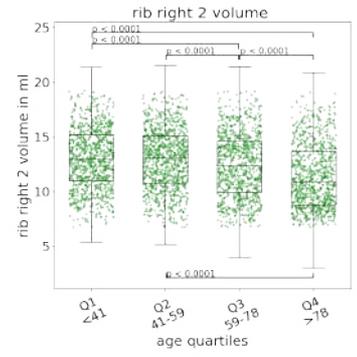
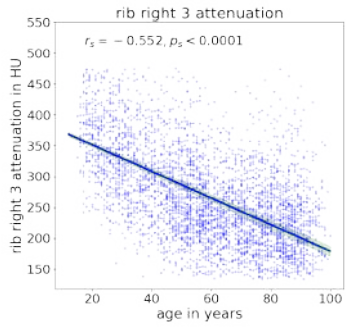 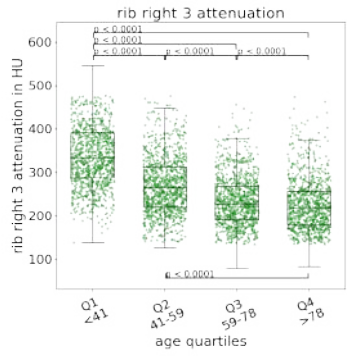 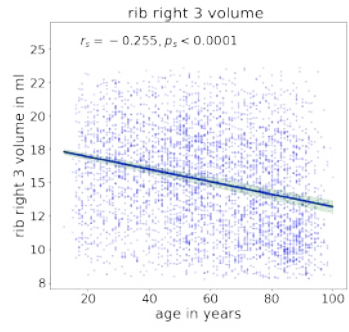 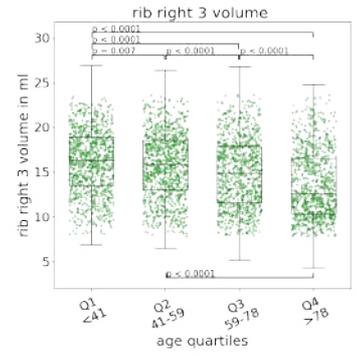
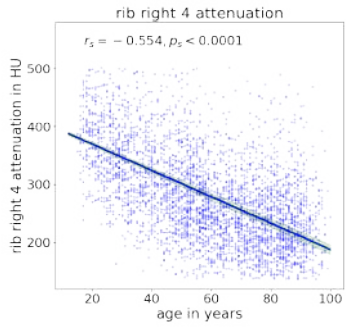 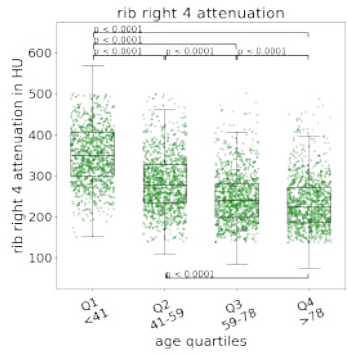 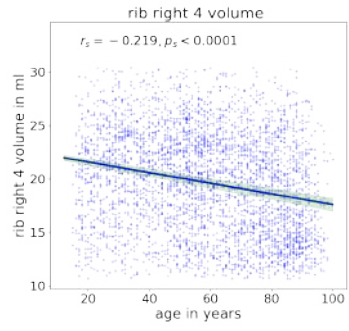 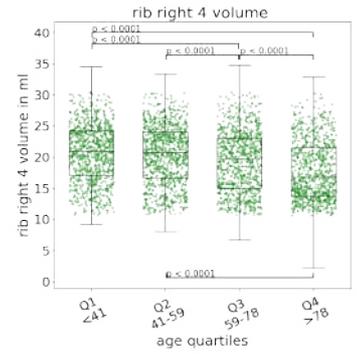

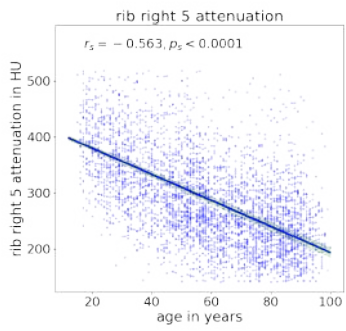 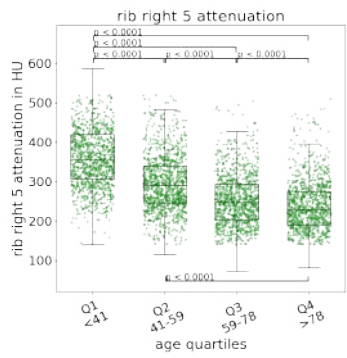 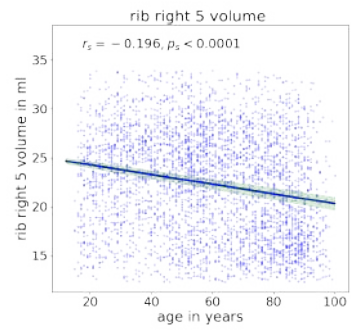 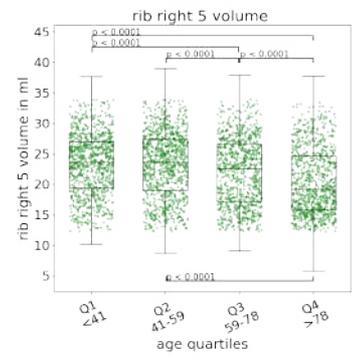
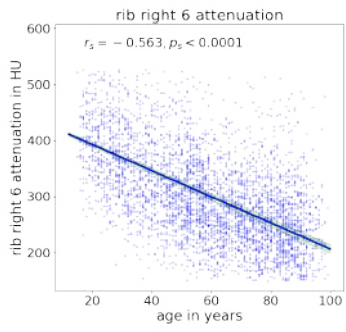 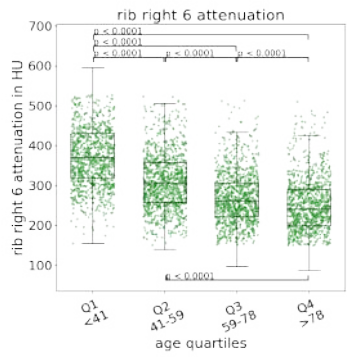 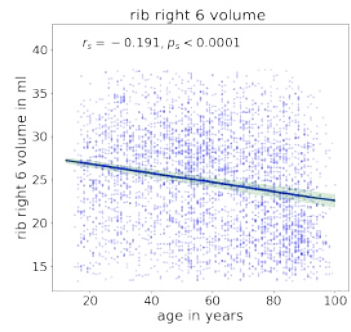 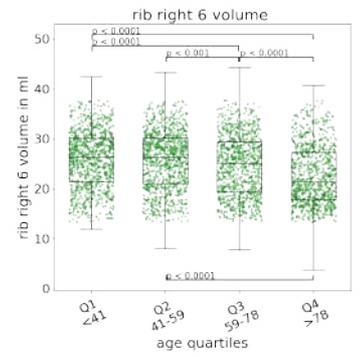
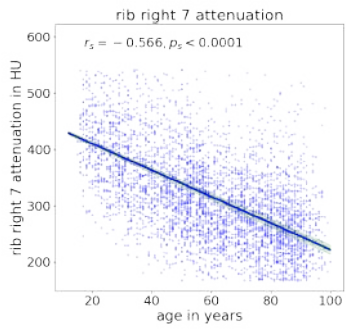 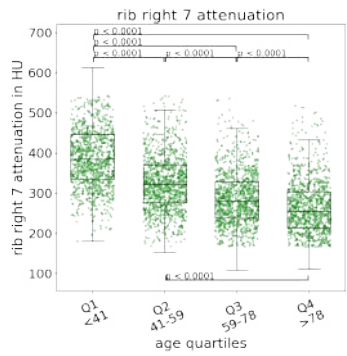 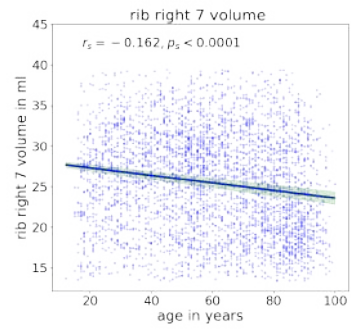 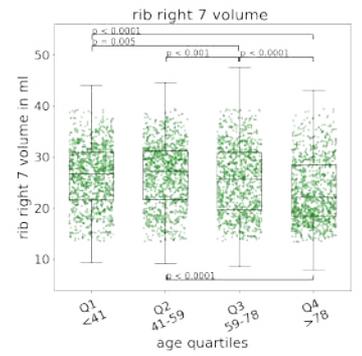
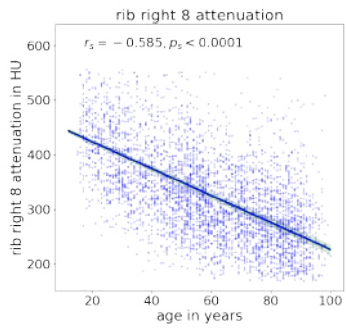 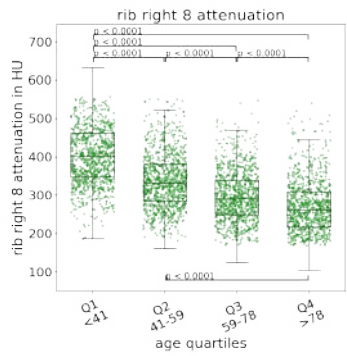 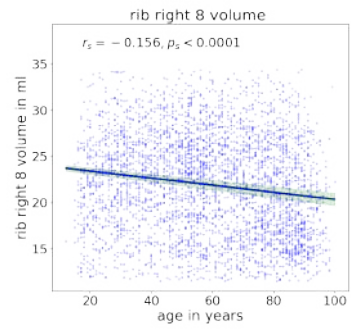 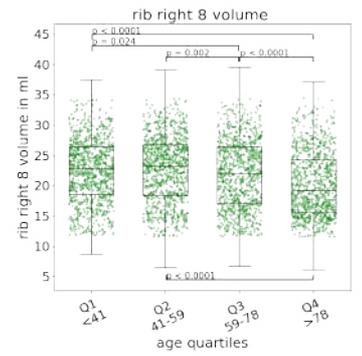
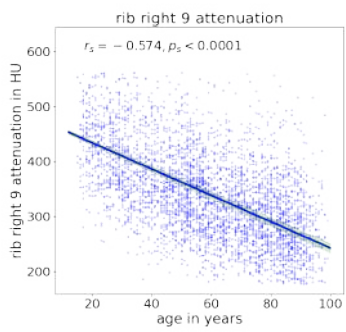 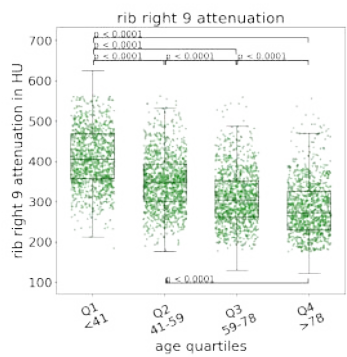 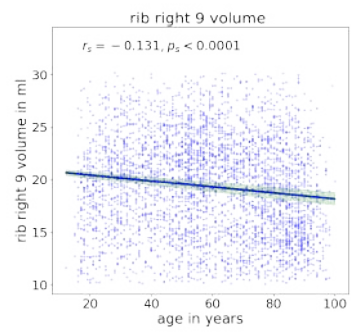 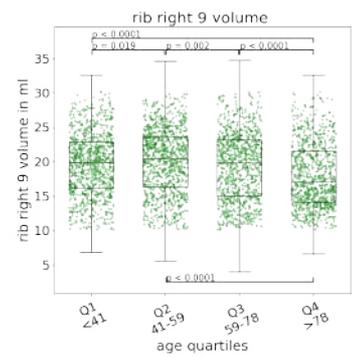

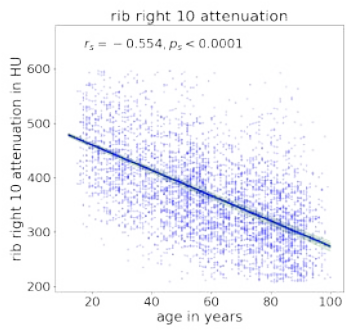 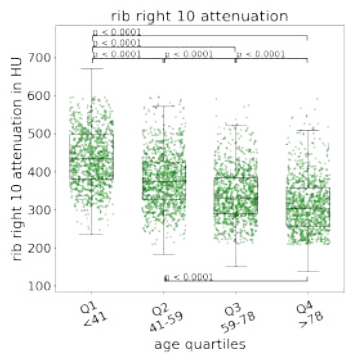 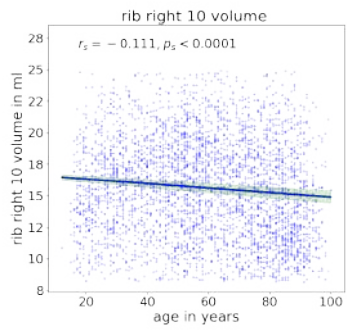 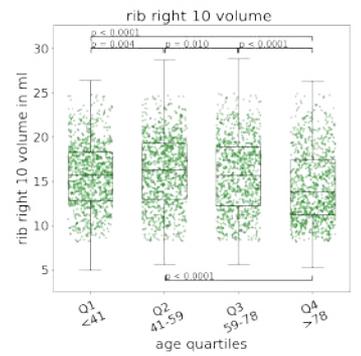
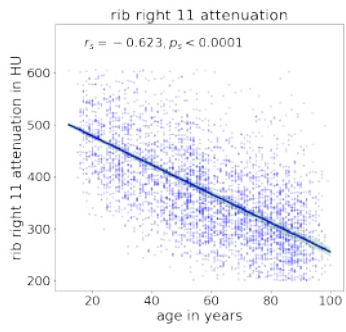 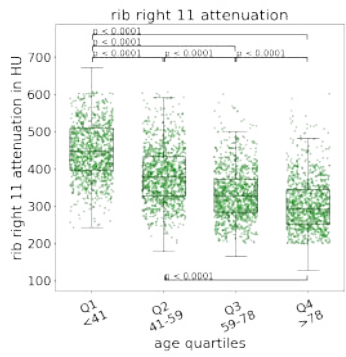 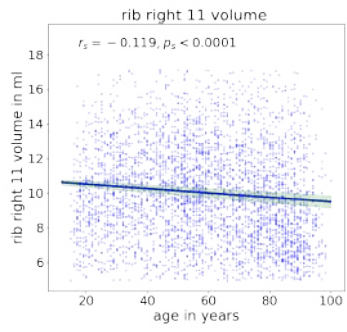 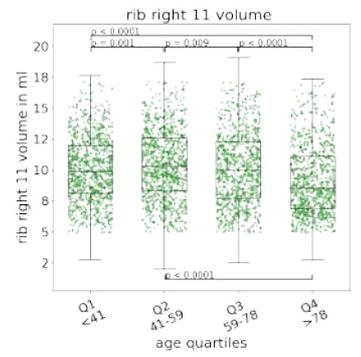
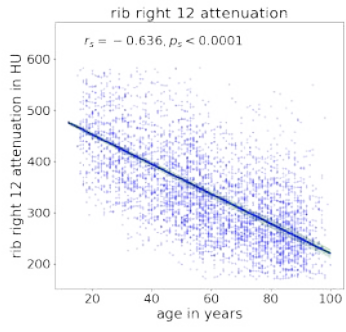 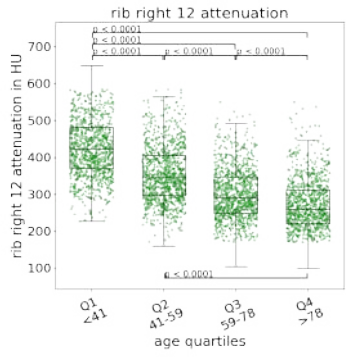 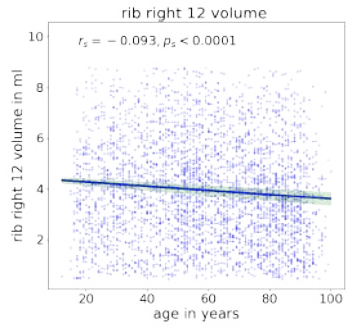 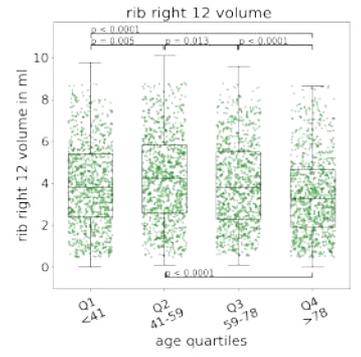
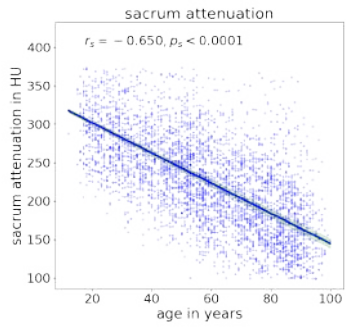 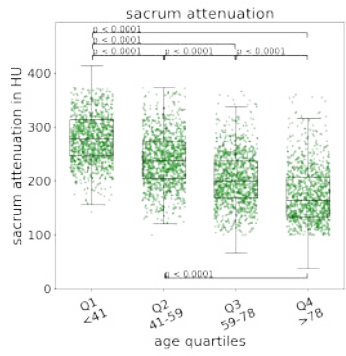 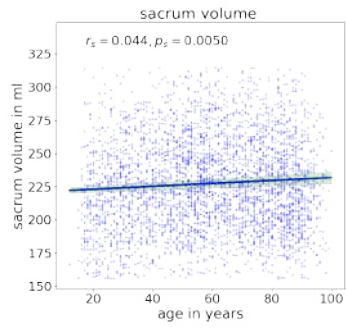 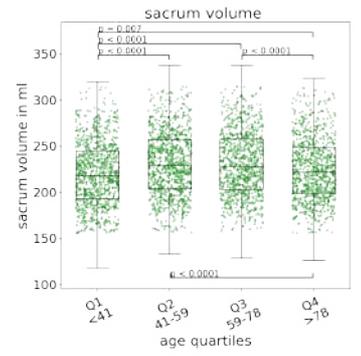
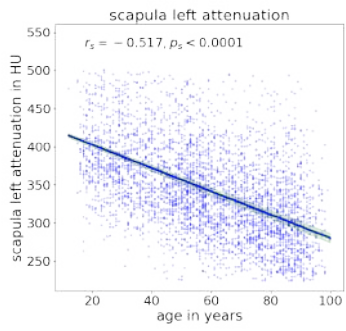 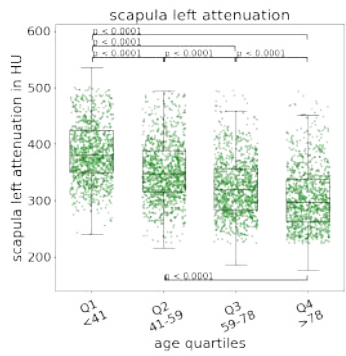 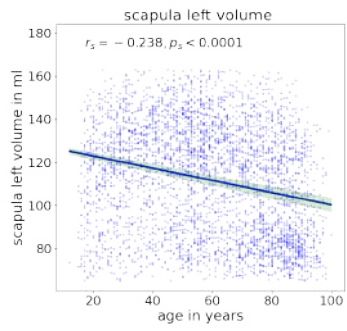 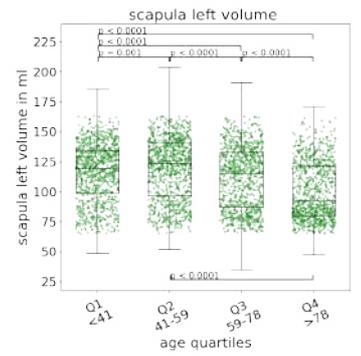

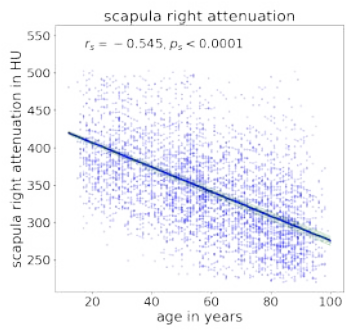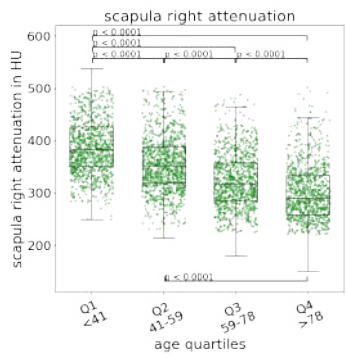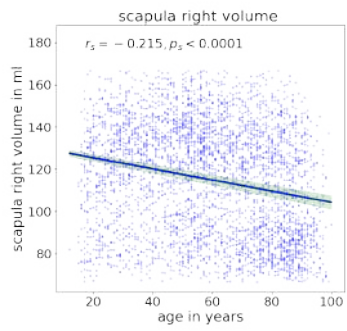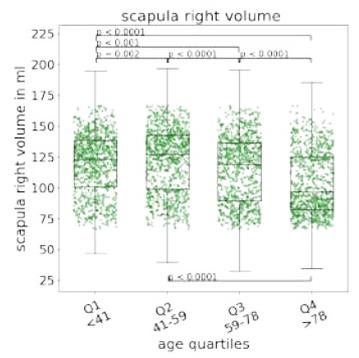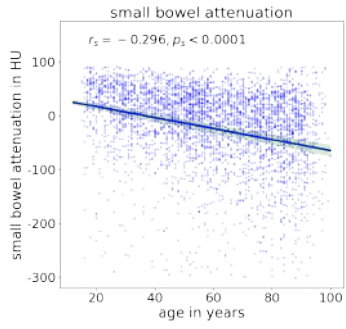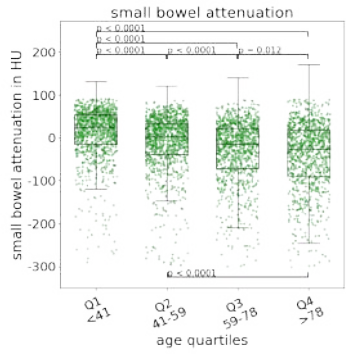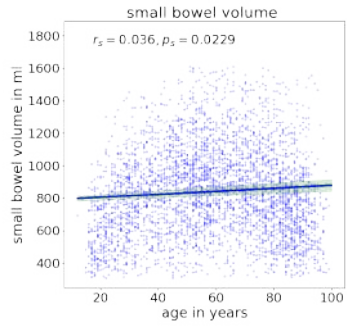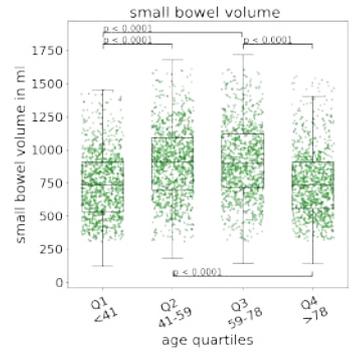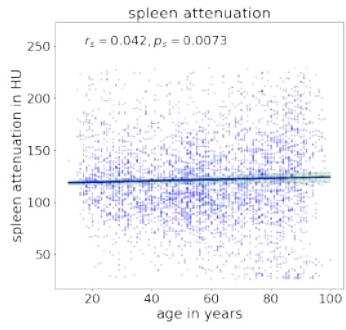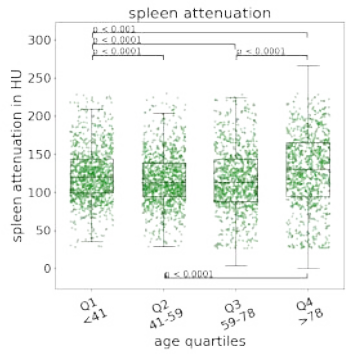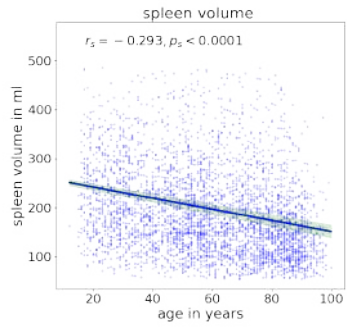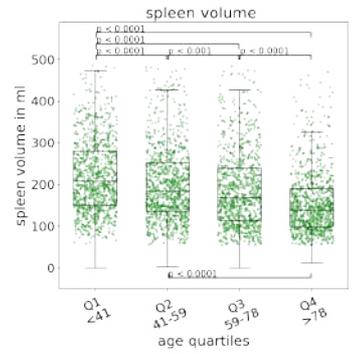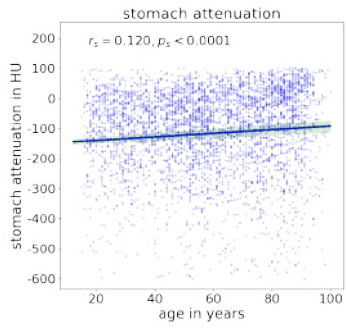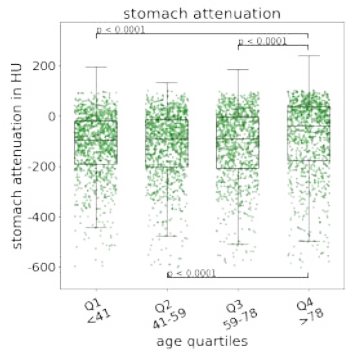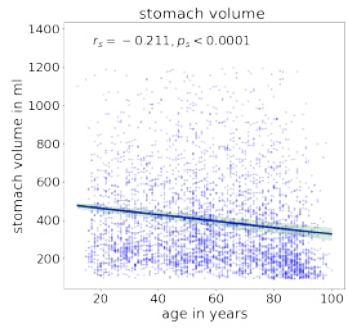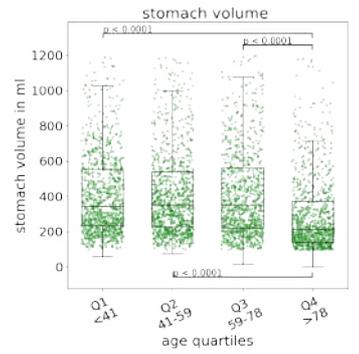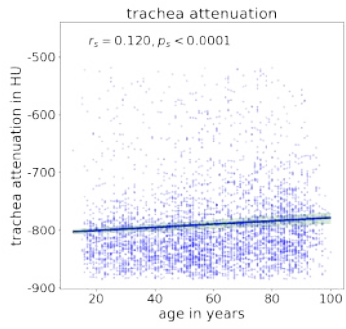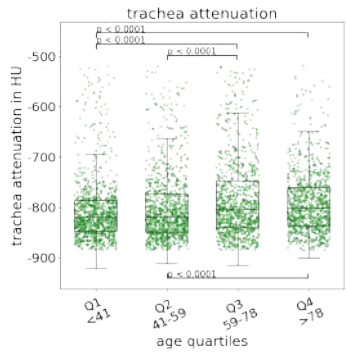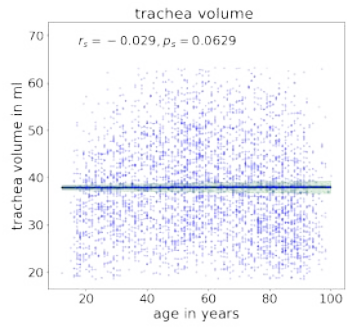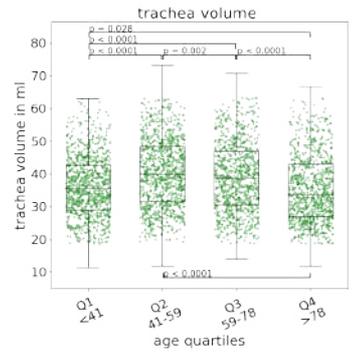

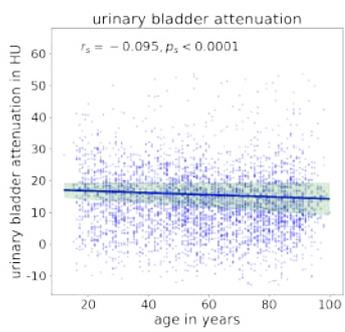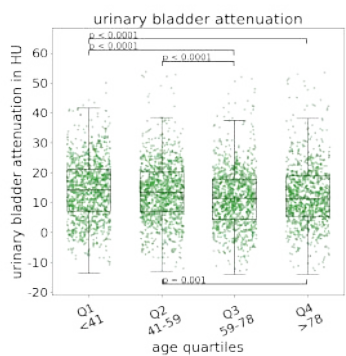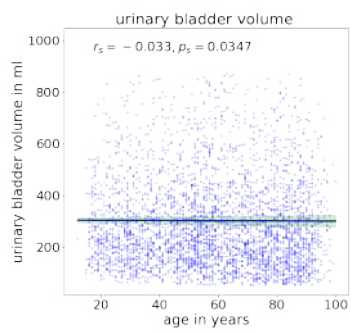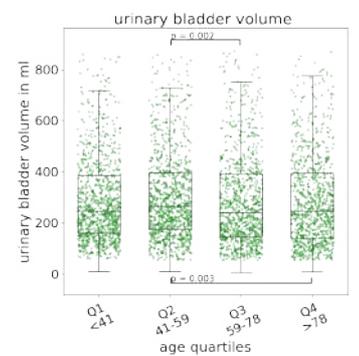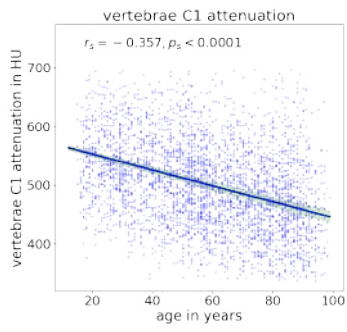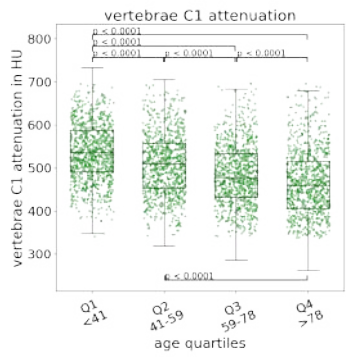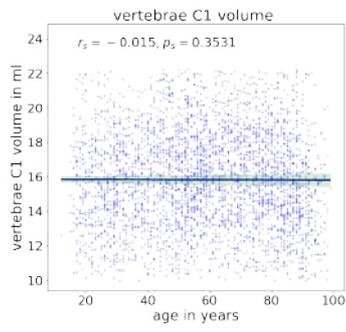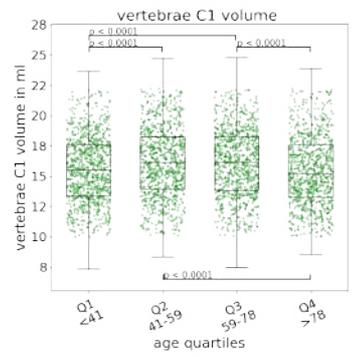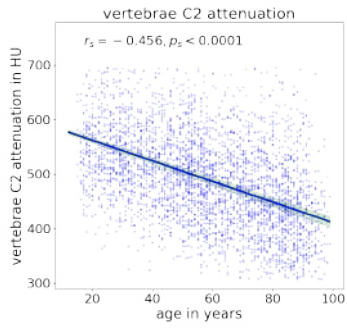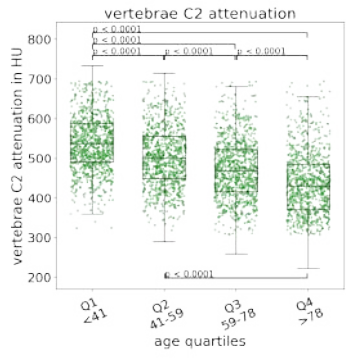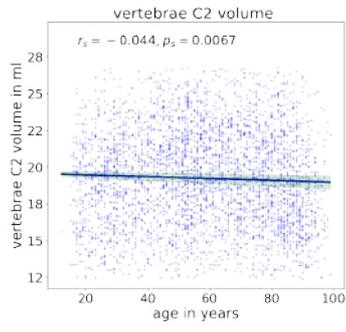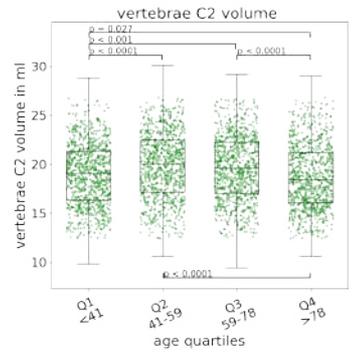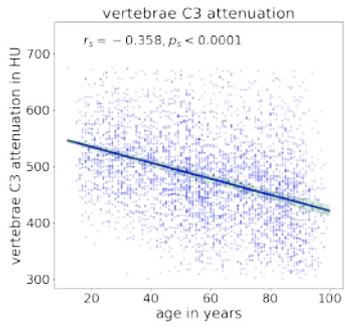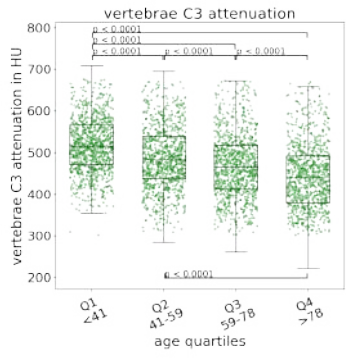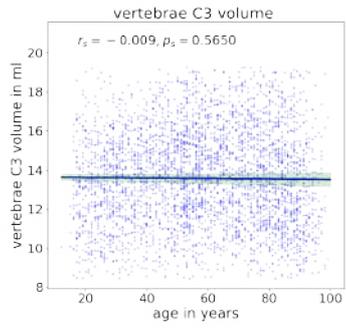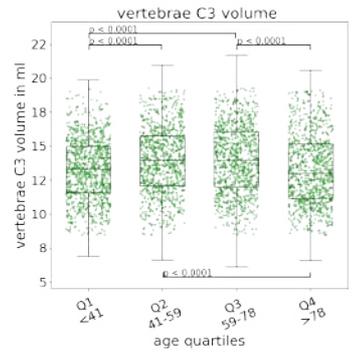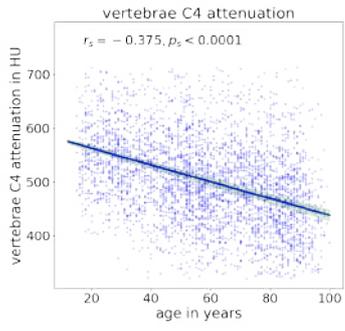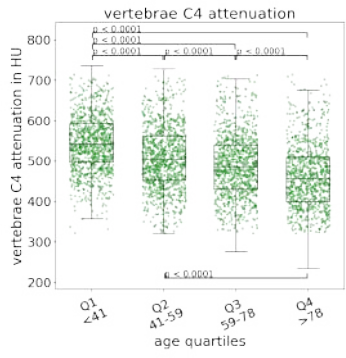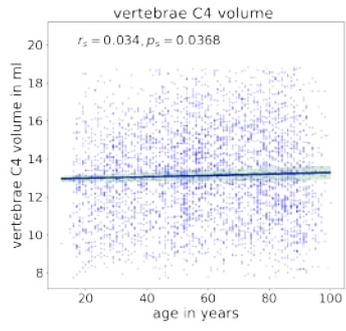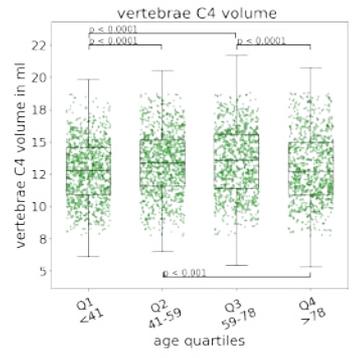

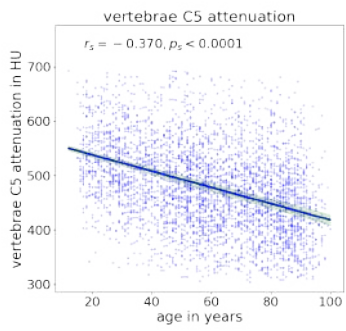 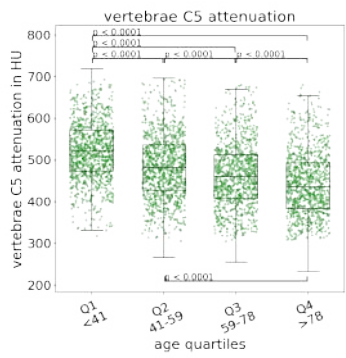 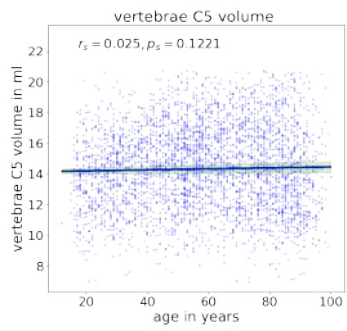 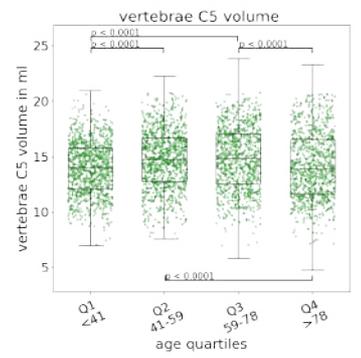
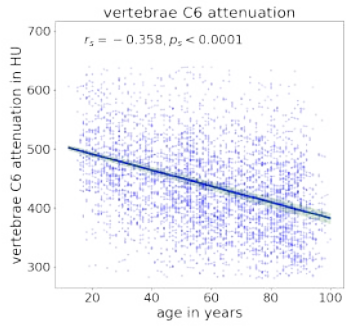 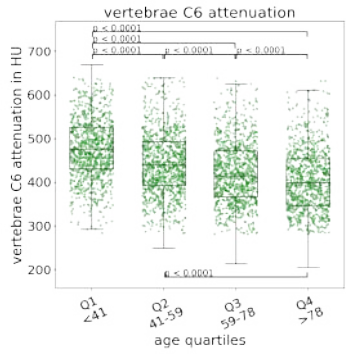 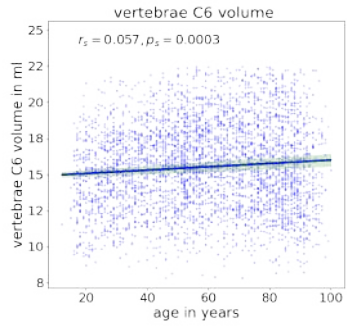 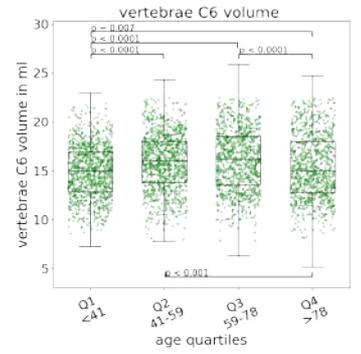
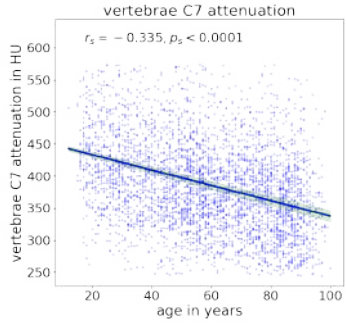 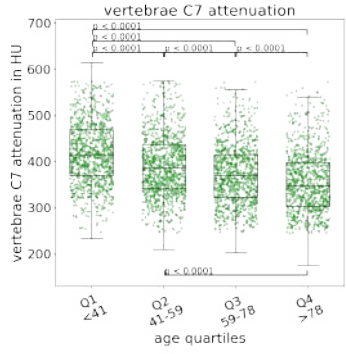 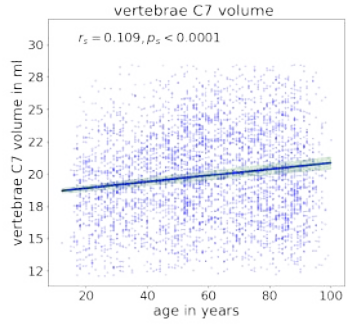 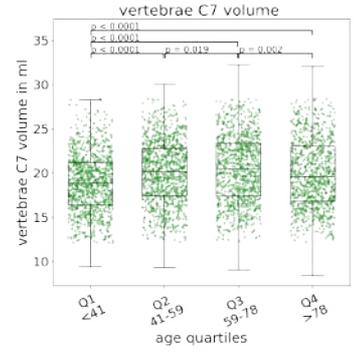
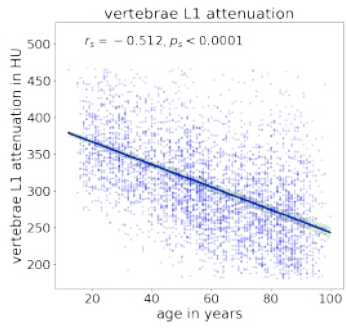 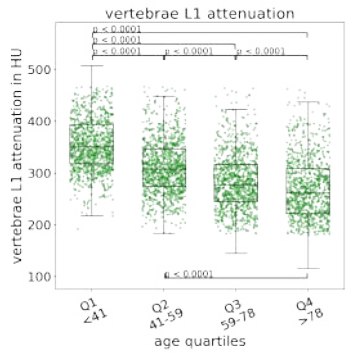 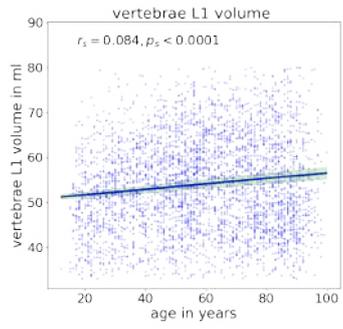 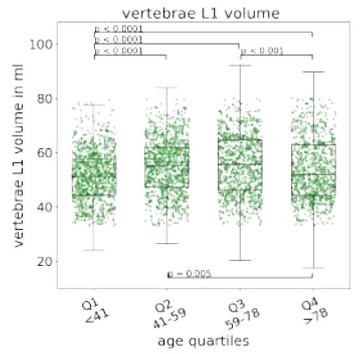
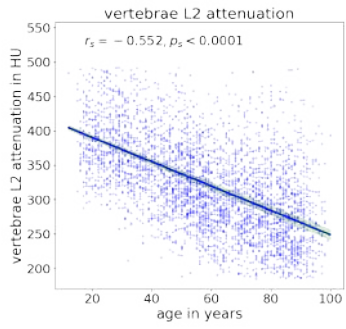 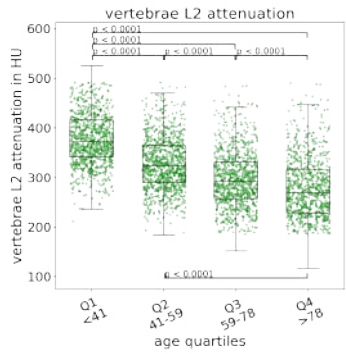 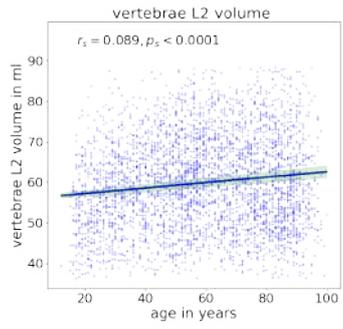 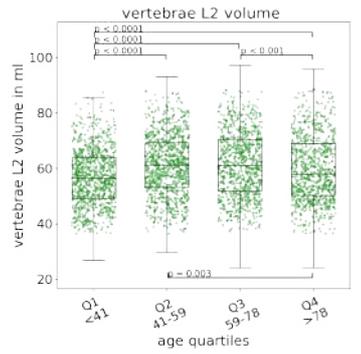

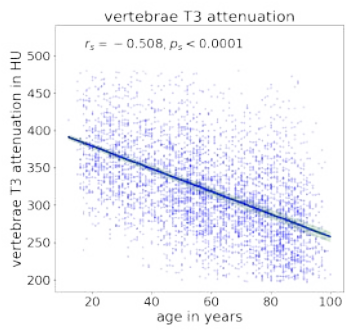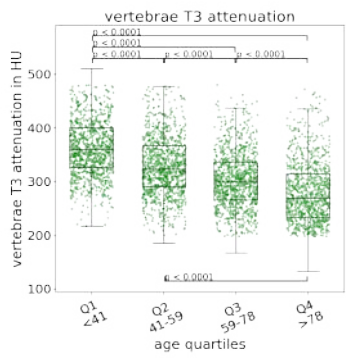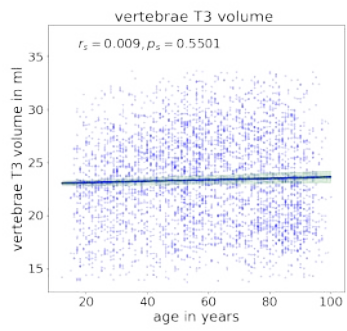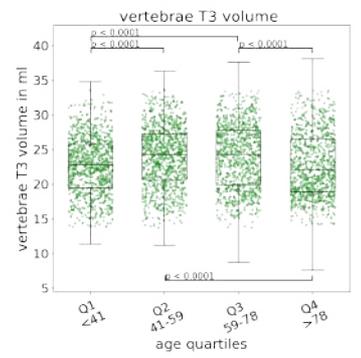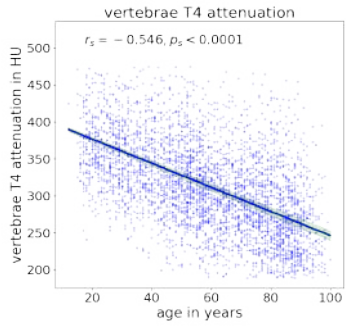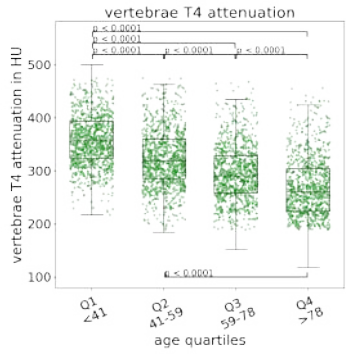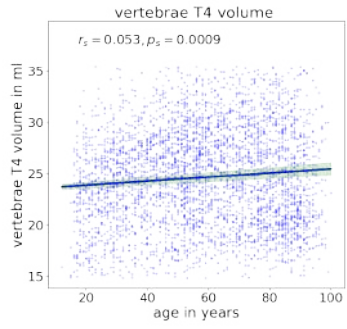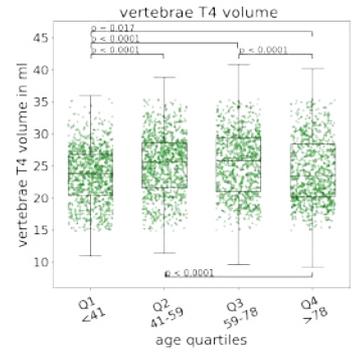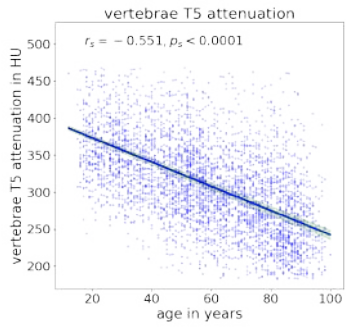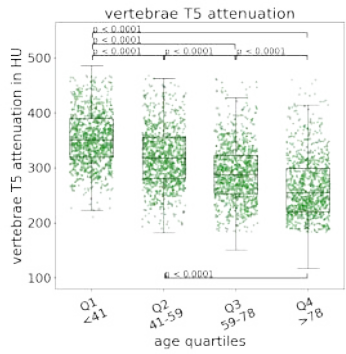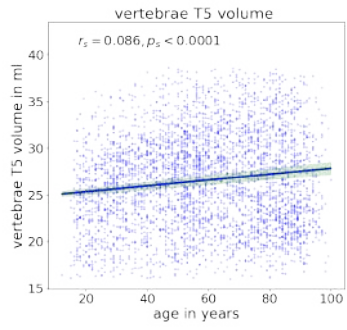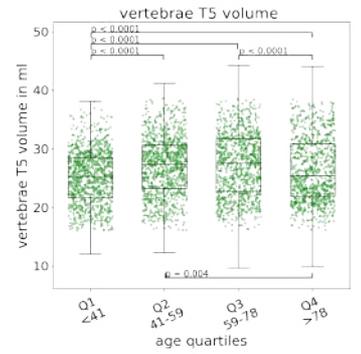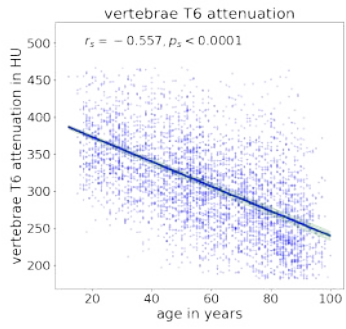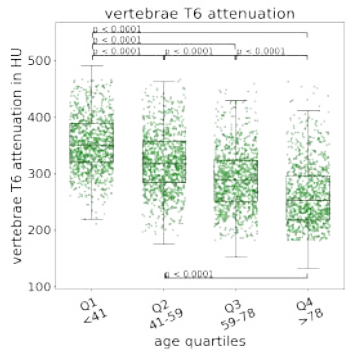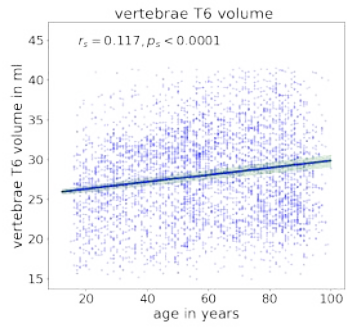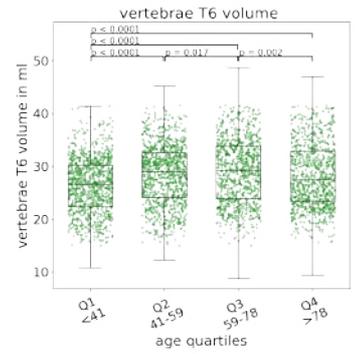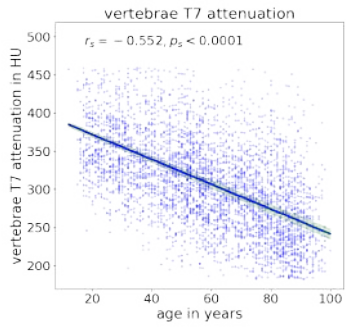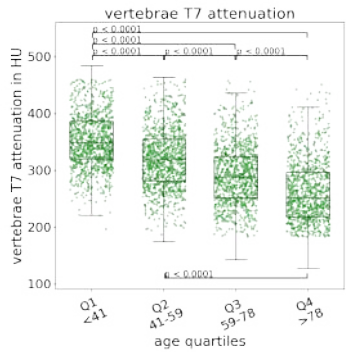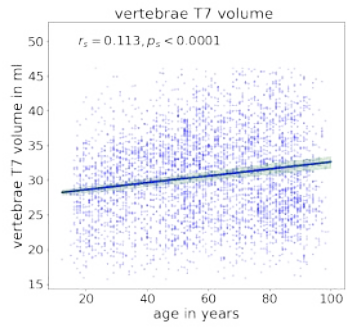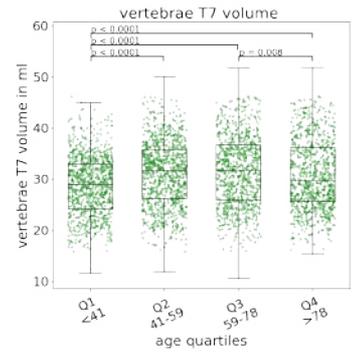

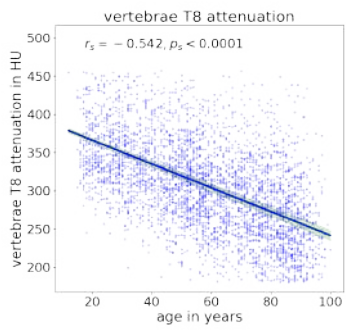 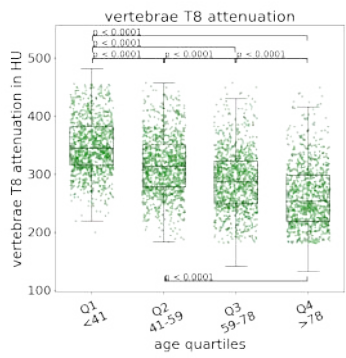 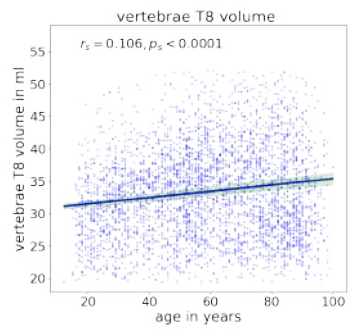 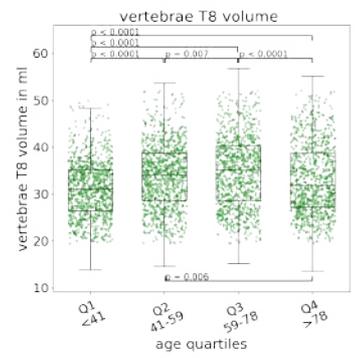
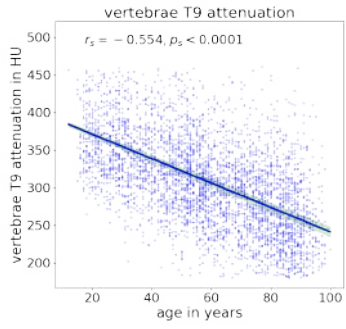 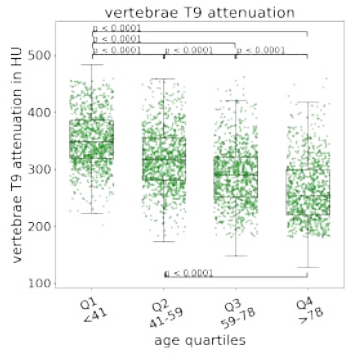 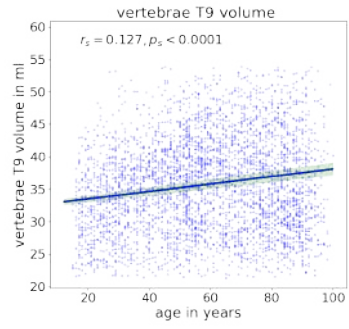 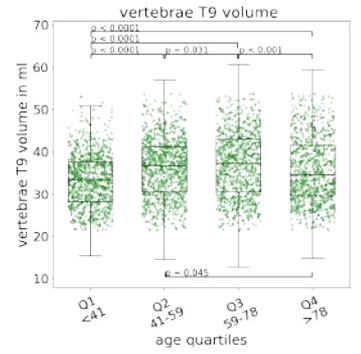
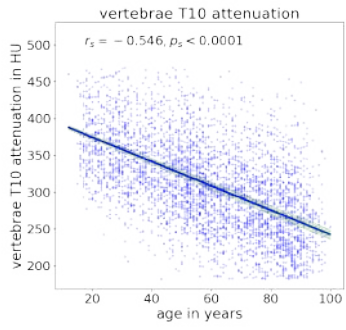 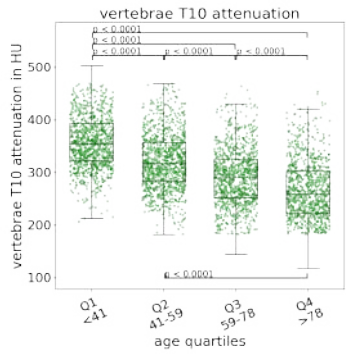 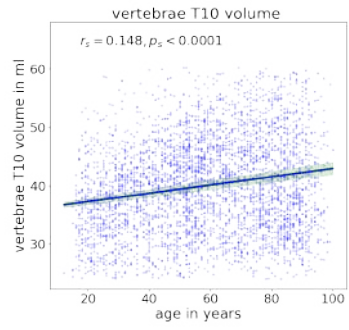 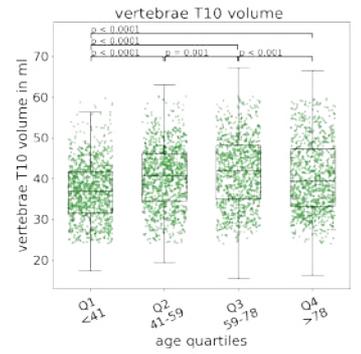
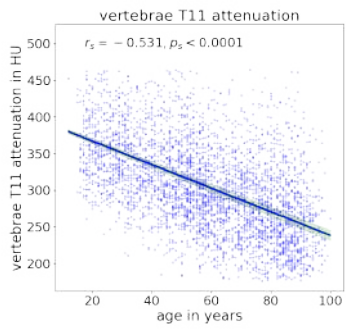 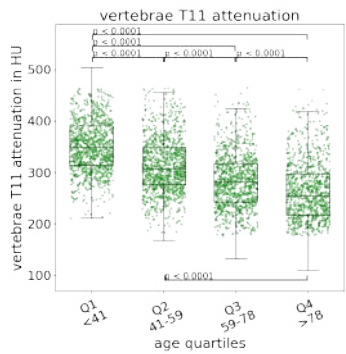 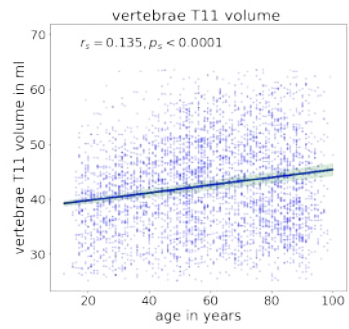 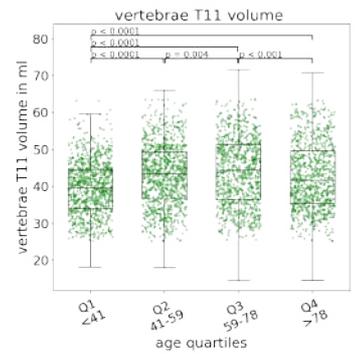
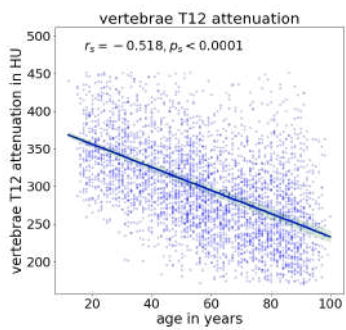 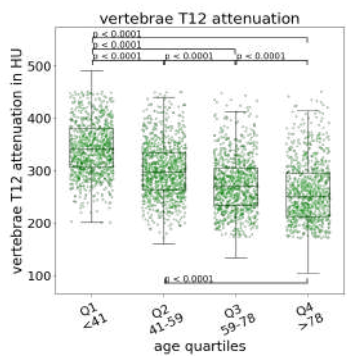 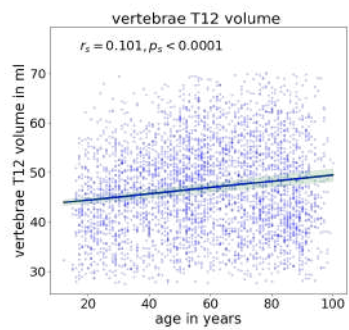 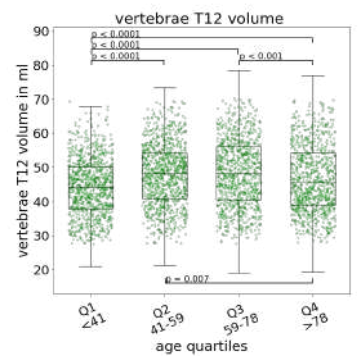

Table S1

P-Value for every body structure regarding volume and attenuation comparing the age quartiles. Group comparison was performed using Kruskal-Wallis test. Significance was reported if the p-value was below 0.0001 due to Bonferroni correction. If a segmentation of a specific body region of a patient was below the lower cut of, the segmentation was excluded.

| Region | Lower Cut Off Volume in ml | Volume p-Value | Volume significant | Density p-Value | Density significant |
|---|---|---|---|---|---|
| spleen | 40 | < 0.0001 | Yes | < 0.0001 | Yes |
| kidney right | 50 | < 0.0001 | Yes | < 0.0001 | Yes |
| kidney left | 30 | < 0.0001 | Yes | < 0.0001 | Yes |
| gallbladder | 1 | < 0.0001 | Yes | < 0.0001 | Yes |
| liver | 100 | < 0.0001 | Yes | < 0.0001 | Yes |
| stomach | 70 | < 0.0001 | Yes | < 0.0001 | Yes |
| aorta | 80 | < 0.0001 | Yes | < 0.0001 | Yes |
| inferior vena cava | 20 | < 0.0001 | Yes | < 0.0001 | Yes |
| portal vein and splenic vein | 1 | < 0.0001 | Yes | < 0.0001 | Yes |
| pancreas | 20 | < 0.0001 | Yes | < 0.0001 | Yes |
| adrenal gland right | 1 | < 0.0001 | Yes | < 0.0001 | Yes |
| adrenal gland left | 1 | < 0.0001 | Yes | < 0.0001 | Yes |
| lung upper lobe left | 100 | < 0.0001 | Yes | < 0.0001 | Yes |
| lung lower lobe left | 100 | < 0.0001 | Yes | 0.2424 | No |
| lung upper lobe right | 100 | < 0.0001 | Yes | < 0.0001 | Yes |
| lung middle lobe right | 100 | < 0.0001 | Yes | < 0.0001 | Yes |
| lung lower lobe right | 100 | < 0.0001 | Yes | 0.2512 | No |
| vertebrae L5 | 40 | < 0.0001 | Yes | < 0.0001 | Yes |
| vertebrae L4 | 40 | < 0.0001 | Yes | < 0.0001 | Yes |
| vertebrae L3 | 30 | < 0.0001 | Yes | < 0.0001 | Yes |
| vertebrae L2 | 30 | < 0.0001 | Yes | < 0.0001 | Yes |
| vertebrae L1 | 30 | < 0.0001 | Yes | < 0.0001 | Yes |
| vertebrae T12 | 20 | < 0.0001 | Yes | < 0.0001 | Yes |
| vertebrae T11 | 20 | < 0.0001 | Yes | < 0.0001 | Yes |
| vertebrae T10 | 20 | < 0.0001 | Yes | < 0.0001 | Yes |
| vertebrae T9 | 20 | < 0.0001 | Yes | < 0.0001 | Yes |
| vertebrae T8 | 20 | < 0.0001 | Yes | < 0.0001 | Yes |
| vertebrae T7 | 10 | < 0.0001 | Yes | < 0.0001 | Yes |
| vertebrae T6 | 10 | < 0.0001 | Yes | < 0.0001 | Yes |
| vertebrae T5 | 10 | < 0.0001 | Yes | < 0.0001 | Yes |
| vertebrae T4 | 10 | < 0.0001 | Yes | < 0.0001 | Yes |
| vertebrae T3 | 10 | < 0.0001 | Yes | < 0.0001 | Yes |
| vertebrae T2 | 10 | < 0.0001 | Yes | < 0.0001 | Yes |
| vertebrae T1 | 10 | < 0.0001 | Yes | < 0.0001 | Yes |
| vertebrae C7 | 10 | < 0.0001 | Yes | < 0.0001 | Yes |
| vertebrae C6 | 1 | < 0.0001 | Yes | < 0.0001 | Yes |
| vertebrae C5 | 1 | < 0.0001 | Yes | < 0.0001 | Yes |
| vertebrae C4 | 1 | < 0.0001 | Yes | < 0.0001 | Yes |
| vertebrae C3 | 1 | < 0.0001 | Yes | < 0.0001 | Yes |
| vertebrae C2 | 1 | < 0.0001 | Yes | < 0.0001 | Yes |
| vertebrae C1 | 1 | < 0.0001 | Yes | < 0.0001 | Yes |
| esophagus | 20 | < 0.0001 | Yes | 0.0010 | No |
| trachea | 20 | < 0.0001 | Yes | < 0.0001 | Yes |
| heart myocardium | 70 | < 0.0001 | Yes | < 0.0001 | Yes |
| heart atrium left | 40 | < 0.0001 | Yes | < 0.0001 | Yes |
| heart ventricle left | 60 | < 0.0001 | Yes | < 0.0001 | Yes |
| heart atrium right | 50 | < 0.0001 | Yes | < 0.0001 | Yes |
| heart ventricle right | 80 | < 0.0001 | Yes | < 0.0001 | Yes |
| pulmonary artery | 30 | < 0.0001 | Yes | < 0.0001 | Yes |
| brain | 1 | < 0.0001 | Yes | < 0.0001 | Yes |
| iliac artery left | 10 | < 0.0001 | Yes | < 0.0001 | Yes |
| iliac artery right | 10 | < 0.0001 | Yes | < 0.0001 | Yes |
| iliac vena left | 10 | < 0.0001 | Yes | < 0.0001 | Yes |
| iliac vena right | 10 | < 0.0001 | Yes | < 0.0001 | Yes |
| small bowel | 100 | < 0.0001 | Yes | < 0.0001 | Yes |
| duodenum | 20 | < 0.0001 | Yes | < 0.0001 | Yes |
| colon | 100 | < 0.0001 | Yes | 0.0006 | No |
| rib left 1 | 1 | < 0.0001 | Yes | < 0.0001 | Yes |
| rib left 2 | 10 | < 0.0001 | Yes | < 0.0001 | Yes |
| rib left 3 | 10 | < 0.0001 | Yes | < 0.0001 | Yes |
| rib left 4 | 10 | < 0.0001 | Yes | < 0.0001 | Yes |
| rib left 5 | 10 | < 0.0001 | Yes | < 0.0001 | Yes |
| rib left 6 | 10 | < 0.0001 | Yes | < 0.0001 | Yes |
| rib left 7 | 10 | < 0.0001 | Yes | < 0.0001 | Yes |
| rib left 8 | 10 | < 0.0001 | Yes | < 0.0001 | Yes |
| rib left 9 | 10 | < 0.0001 | Yes | < 0.0001 | Yes |
| rib left 10 | 10 | < 0.0001 | Yes | < 0.0001 | Yes |
| rib left 11 | 1 | < 0.0001 | Yes | < 0.0001 | Yes |
| rib left 12 | 1 | < 0.0001 | Yes | < 0.0001 | Yes |
| rib right 1 | 1 | < 0.0001 | Yes | < 0.0001 | Yes |
| rib right 2 | 10 | < 0.0001 | Yes | < 0.0001 | Yes |
| rib right 3 | 10 | < 0.0001 | Yes | < 0.0001 | Yes |
| rib right 4 | 10 | < 0.0001 | Yes | < 0.0001 | Yes |
| rib right 5 | 10 | < 0.0001 | Yes | < 0.0001 | Yes |
| rib right 6 | 10 | < 0.0001 | Yes | < 0.0001 | Yes |
| rib right 7 | 10 | < 0.0001 | Yes | < 0.0001 | Yes |
| rib right 8 | 10 | < 0.0001 | Yes | < 0.0001 | Yes |
| rib right 9 | 10 | < 0.0001 | Yes | < 0.0001 | Yes |
| rib right 10 | 10 | < 0.0001 | Yes | < 0.0001 | Yes |
| rib right 11 | 1 | < 0.0001 | Yes | < 0.0001 | Yes |
| rib right 12 | 1 | < 0.0001 | Yes | < 0.0001 | Yes |
| humerus left | 30 | < 0.0001 | Yes | < 0.0001 | Yes |
| humerus right | 20 | < 0.0001 | Yes | < 0.0001 | Yes |
| scapula left | 60 | < 0.0001 | Yes | < 0.0001 | Yes |
| scapula right | 60 | < 0.0001 | Yes | < 0.0001 | Yes |
| clavicula left | 20 | < 0.0001 | Yes | < 0.0001 | Yes |
| clavicula right | 20 | < 0.0001 | Yes | < 0.0001 | Yes |
| femur left | 90 | < 0.0001 | Yes | < 0.0001 | Yes |
| femur right | 90 | < 0.0001 | Yes | < 0.0001 | Yes |
| hip left | 100 | < 0.0001 | Yes | < 0.0001 | Yes |
| hip right | 100 | < 0.0001 | Yes | < 0.0001 | Yes |
| sacrum | 100 | < 0.0001 | Yes | < 0.0001 | Yes |
| face | 1 | < 0.0001 | Yes | < 0.0001 | Yes |
| gluteus maximus left | 100 | < 0.0001 | Yes | < 0.0001 | Yes |
| gluteus maximus right | 100 | < 0.0001 | Yes | < 0.0001 | Yes |
| gluteus medius left | 100 | < 0.0001 | Yes | < 0.0001 | Yes |
| gluteus medius right | 100 | < 0.0001 | Yes | < 0.0001 | Yes |
| gluteus minimus left | 30 | < 0.0001 | Yes | < 0.0001 | Yes |
| gluteus minimus right | 30 | < 0.0001 | Yes | < 0.0001 | Yes |
| autochthon left | 100 | < 0.0001 | Yes | < 0.0001 | Yes |
| autochthon right | 100 | < 0.0001 | Yes | < 0.0001 | Yes |
| iliopsoas left | 100 | < 0.0001 | Yes | < 0.0001 | Yes |
| iliopsoas right | 100 | < 0.0001 | Yes | < 0.0001 | Yes |
| urinary bladder | 40 | 0.0041 | No | < 0.0001 | Yes |